\documentstyle[aps,prb,preprint]{revtex}
\begin{document}
\title{The Ruderman-Kittel-Kasuya-Yosida (RKKY) interaction across a
tunneling junction out of equilibrium}
\draft
\author{N.\ F.\ Schwabe and R.\ J.\ Elliott}
\address{University of Oxford,  Department of Physics,\\
Theoretical Physics, 1 Keble Road, Oxford OX1 3NP, United Kingdom}

\author{Ned S.\ Wingreen}
\address{NEC Research Institute, 4 Independence Way, Princeton NJ
08540, USA}

\date{\today}
\maketitle

\vspace{-7mm}
\begin{abstract}
\vspace{-10mm}

The Ruderman-Kittel-Kasuya-Yosida (RKKY) interaction between two
magnetic 
$s$-$d$ spin impurities across a tunneling
junction is studied when the system is driven out of equilibrium
through biasing the junction. The nonequilibrium situation is handled
with the Keldysh time-loop perturbation formalism in conjunction with
 appropriate coupling methods
for tunneling systems due to Caroli and Feuchtwang. 
We find that the presence of a nonequilibrium bias across the junction
leads to an interference of several fundamental oscillations, such
that in this tunneling geometry, it is possible 
to tune the interaction
between ferromagnetic and antiferromagnetic coupling at a fixed
impurity configuration, simply by changing the bias across the
junction. Furthermore, it is shown that the range of the RKKY
interaction 
is altered out of equilibrium, such that in particular the
interaction energy between two slabs of spins scales
extensively with the thickness of the slabs in the presence of an
applied bias. 
\end{abstract}
\vspace{1ex}
\section{Introduction}

The Ruderman-Kittel-Kasuya-Yosida (RKKY) interaction 
has been a very intensely studied phenomenon in solid state
 physics since it was first
proposed as an interaction between nuclear spins\cite{bi0,bi01}, and
later 
between localized electronic spins in metals\cite{bi02}.
More recently studies of the interaction have focussed on its effects
in various structured systems and in particular on
the role it plays in the giant magnetoresistance effects observed in
some layered  structures of magnetic and non-magnetic
materials\cite{bi4,gm1,gm2,gm3}.

Furthermore it has been suggested that the RKKY interaction is
responsible for spin polarization effects observed in tunneling
systems\cite{bi3}, such as layered structures with a potential
barrier formed by an insulating oxide layer or a vacuum gap.
For arrangements involving movable (vacuum) tunneling junctions, such
as those occurring in scanning tunneling microscopy (STM) it has been
suggested that the exchange interaction between two magnetic
materials on either side of the tunneling junction can be used 
in a modified version of atomic force microscopy, which may be called 
exchange force microscopy, in order 
to resolve an atomic image of a sample structure\cite{bi6,bi21}. 

Particularly in tunneling systems, the measurement of electronic
properties, such as for example the electronic density of states,
intrinsically requires the structure to be biased out of
equilibrium. The presence of a nonequilibrium bias may also occur in
structures which exhibit giant magnetoresistance phenomena, when parts
of the structure contain a potential barrier.
To date, however, descriptions of 
the RKKY interaction are confined to systems in equilibrium, including
approximate theoretical treatments of the interaction across a
potential well\cite{bi4} and a tunneling barrier\cite{bi21}. 

It is therefore the purpose of the present paper to establish a
theoretical description of the RKKY-interaction across a tunneling
barrier out of equilibrium from first principles. For this purpose we
employ the Keldysh nonequilibrium perturbation formalism\cite{bi12}
along with a 
coupling procedure for structured systems developed by 
Caroli, Combescot {\it et al.}\cite{bi13,bi14,bi14_5,bi15} and
Feuchtwang\cite{bi16,bi17,bi18}, leading 
to a proper nonequilibrium field theoretic 
description.  
Since the treatments by Caroli and Feuchtwang are proper many-body
formalisms their application to the  
present problem 
simultaneously provides the basis for the inclusion of further
many-body effects such as carrier-carrier
 or carrier-phonon interactions to
the problem. 

With this technique we manage to derive a general nonequilibrium
solution for the RKKY interaction  for
various dimensionalities and arrangements of spins
within a structured system. Particularly we show that the interaction
of spins across a tunneling junction can be tuned between
ferromagnetic 
(FM) and antiferromagnetic (AFM) coupling by changing the bias 
across the junction alone. For the application to exchange force
microscopy, mentioned before,
 it is shown that varying the bias across the vacuum junction 
in an STM can lead to a force which
switches between attractive and repulsive behavior and that this force
should be experimentally measurable with a state of the art apparatus.
Furthermore, we show that the presence of a nonequilibrium bias across
the junction significantly alters the range of the RKKY interaction,
 such that in particular the
interaction energy between two slabs of spins scales
extensively with the thickness of the slabs.

The remainder of the paper is structured as follows: In Sec.\ II we
establish a model of the tunneling system containing the spin
impurities. Sec.\
III contains a derivation 
of the RKKY interaction out of equilibrium in various dimensions in
terms of general Keldysh nonequilibrium Green's functions. In Sec.\
III A we consider the interaction of two magnetic $s$-$d$ impurities
across a one dimensional  tunneling junction out of equilibrium and in
Sec.\ III B  we
extend these results to the more realistic situation
of two magnetic layers or slabs of spins on either side of a
planar three dimensional tunneling junction, including the possibility
of a non-magnetic spacer material between the magnetic materials and
the barrier. Sec.\ IV
describes how the Keldysh Green's functions used in Sec.\ III are
obtained in a system 
containing a potential barrier. In Sec.\ V we implement our results
numerically and calculate equilibrium and nonequilibrium versions of
the RKKY interaction for various tunneling geometries. Sec.\ VI 
discusses the experimental observability of the behavior of the
interaction 
predicted by the numerical study. In Sec.\ VII, in conclusion,
the implications and possible further applications
of this work are summarized.

\section{The model}

The one dimensional tunneling system we consider consists of
a tunneling barrier which is connected to two leads on the left and
right at the points $L$ and $R$, respectively, as shown in Fig.\ 1. In
equilibrium the barrier is assumed to be flat on top with an abrupt 
potential change of $V_0$ at the interfaces. The corresponding single
particle 
potential can be written as
\begin{equation}
V(x)=V_0 \Theta (x-L) \Theta (R-x). \label{mv}
\end{equation}

Upon biasing the junction, the potential of the barrier acquires a
slope and the chemical potential $\mu^R$ in the right lead 
undergoes a shift $eV$ with respect to the chemical
potential $\mu^L$ in the left lead. 
Simultaneously, the conduction band bottom $V^R$
in the right lead shifts by $eV$ with respect to
conduction band bottom $V^L$ in the left lead,
such that the electronic potential is changed to
\begin{equation}
V(x)=\left[ V_0  - eV \frac {x-L} {R-L} \right] \Theta (x-L) \Theta
(R-x) - eV \Theta (x-R). \label{nv}
\end{equation}
Here the bottom of the conduction band on the left side is taken as
the origin of energy, $e$ is the modulus of the elementary charge of
an electron, and $V$ the voltage drop across the barrier. 

Two magnetic $s$-$d$ impurities
 are situated within the electrodes, on either side of the barrier, at
an equal distance $d$ from the interfaces.  For simplicity, both
electrodes are 
considered to consist of the same material and the effective electron
masses are assumed to be equal in all three parts of the junction.

The Hamiltonian of the system, with a single electron band in each
part of the junction, can be written as
\begin{eqnarray}
H & = & \sum_{\alpha} \int dx \Psi_{\alpha}^{\dagger}(x)\left[ \frac
{p^2} {2m} + V(x) \right] \Psi_{\alpha}(x) \nonumber \\
& - & \frac J 2 \sum_{p=1,2;\alpha,\beta} {\mbox {\boldmath
$\sigma$}}_{\alpha \beta} \cdot {\bf S}_p \int dx
\Psi_{\alpha}^{\dagger}(x) 
\delta(x-x_p) \Psi_{\beta}(x) = H_0 + H_{s{\rm -}d} , \label{m1}
\end{eqnarray}
where $H_{s{\rm -}d}$ is the last, spin dependent, term in the
Hamiltonian and the coupling constant $J$ is assumed to have the units
${\rm Jm}^d$, where $d$ is the dimensionality of the system
considered. Here a space representation has been chosen, which proves
to be advantageous due to the lack of translational invariance caused
by the presence of the barrier potential $V(x)$. $p^2/2m$ is the
kinetic energy of electrons with uniform effective mass $m$, and we
use $\hbar \equiv 1$ throughout. ${\mbox {\boldmath
$\sigma$}}_{\alpha\beta}$ is the vector of 
Pauli matrices, used to represent the spin of the conduction electrons
which couple to the two local moments ${\bf S}_p$. The indices
$\alpha$ and $\beta$ label the two spin components: $\alpha, \beta \in
\{ \uparrow , \downarrow \}$. 

\section{Expression for the RKKY interaction}

\subsection{The interaction in one dimension}

The RKKY interaction energy between
the two impurities is  
calculated from the lowest order exchange contributions 
to the perturbation of the energy of the localized spins in the
presence of the conduction electrons. Either in or out of equilibrium
this energy is given by the expectation value $\langle H_{s{\rm  
-}d} \rangle_{\rm (1ex)}$, where the subscript (1ex) indicates that
only the first order exchange contributions to the average are
considered. Effectively, therefore, one has to calculate the first  
order perturbation to the conduction electron spin density due to the
first spin at the 
location of the second spin and {\it vice versa}\cite{bi0,bi1}. 
The total interaction energy is a sum of these two contributions
\begin{equation}
E_{\rm RK}= - \frac J 2 \sum_{p=1,2} {\bf S}_p \cdot \Delta
{\mbox {\boldmath $\rho$}}_{\rm (1ex)}(x_p) , \label{i1}
\end{equation}
where
\begin{equation}
\Delta {\mbox {\boldmath $\rho$}}_{\rm (1ex)} (x_p)=
\sum_{\alpha,\beta} {\mbox {\boldmath $\sigma$}}_{\alpha\beta} \langle
n_{\alpha \beta}(x_p) \rangle_{\rm (1ex)} =
\sum_{\alpha,\beta} {\mbox {\boldmath $\sigma$}}_{\alpha\beta}
\langle \Psi^{\dagger}_{\alpha}(x_p) \Psi_{\beta}(x_p) 
\rangle_{\rm (1ex)}  \hspace{5mm} (p \in \{1,2\}) \label{i2}
\end{equation}
is the first order exchange contribution to the spin density at site
$p$. 

In the present case we now have to calculate (\ref{i1}) by means of
(\ref{i2}) out of equilibrium. In order to do so we make use of the
Keldysh formalism.
In the Keldysh notation the spin dependent particle
density correlation function $\langle n_{\alpha\beta}(x) \rangle =
\langle 
\Psi^{\dagger}_{\alpha}(x) \Psi_{\beta}(x) \rangle$ can  be written in
terms of the 
Keldysh Green's function $G_{\alpha\beta}^<(x,t;x',0)$ which is
defined as 
\begin{equation}
G_{\alpha\beta}^<(x,t;x',0)= i \langle \Psi_{\beta}^{\dagger}(x',0)
\Psi_{\alpha}(x,t) \rangle,
 \label{i3} 
\end{equation}
where the operators $\Psi^{\dagger}_{\alpha}(x,t)$ and
$\Psi_{\alpha}(x,t)$ 
are the field operators of the system considered, which create and
destroy a particle with spin $\alpha$ at point $x$ and time $t$,
respectively. 
From this we find
\begin{equation}
\langle n_{\alpha\beta}(x) \rangle = -i \lim_{\stackrel {\scriptstyle
t \rightarrow 0}{x' \rightarrow x }}
\int_{-\infty}^{\infty} \frac {d \omega} {2 \pi} e^{i\omega
t}G_{\alpha\beta}^<(x,x';\omega). \label{i4} 
\end{equation}
Using the definition of
$G^<$, it can be shown in equilibrium that this expression is
equivalent to the usual relation between the particle density and the 
 retarded Green's function $G^r$,
\begin{equation}
\langle n_{\alpha\beta}(x) \rangle = - \frac 1 {\pi} \lim_{x'
\rightarrow x } 
\int_{-\infty}^{\infty} n_F(\omega) {\rm
Im} \left[ G_{\alpha\beta}^r(x,x';\omega)\right] d \omega ,
\label{i4_5} 
\end{equation}
where $n_F(\omega)= \{ {\rm exp} [ \beta(\omega - \mu) ] +1 \}^{-1} $
is the 
usual Fermi distribution function in equilibrium, $\mu$ is the
equilibrium chemical potential and $G^r$ is the
retarded Green's function of the system, which is 
defined as 
\begin{equation}
G^r_{\alpha \beta} (x,x';t) =  -i\Theta(t)\langle \{
\Psi_{\beta}^{\dagger}(x',0),\Psi_{\alpha}(x,t) \} \rangle ,
\end{equation}
 where $\{$ , $\}$ denotes the anti-commutator.
The advanced Green's function $G^a$
\begin{equation}
G^a_{\alpha \beta}(x,x';t) = i \Theta(-t)\langle \{
\Psi_{\beta}^{\dagger}(x',0),\Psi_{\alpha}(x,t) \} \rangle ,
\end{equation} 
will be used later.

In order to obtain a proper nonequilibrium result for the
RKKY interaction we have to perform first order nonequilibrium
perturbation theory on expression (\ref{i4}). The appropriate
formalism, introduced by Keldysh, reformulates the regular
diagrammatic 
perturbation theory in terms of a 2 $\times$ 2 matrix formalism, to
properly handle the time development, since out of equilibrium the
usual 
$S$-matrix expansion based on the Gell-Mann-Low theorem breaks
down\cite{bi12,bim}. The Green's function matrix of the unperturbed
system without 
spin impurities, but including the full barrier potential within this
formalism is written as
\begin{equation}
{\bf G}_{(0) \alpha\beta}(x,x';t)= \left( {\begin{array}{cc}
G^t_{(0)\alpha\beta}(x,x';t) & - G_{(0)\alpha\beta}^<(x,x';t) \\
G^>_{(0)\alpha\beta}(x,x';t) & - G_{(0)\alpha\beta}^{\tilde t}(x,x';t)
\end{array}} \right) \label{i5_5}.
\end{equation}
$G^t$ and $G^{\tilde t}$ denote the usual time-ordered and
anti-time-ordered  
Green's functions with the general definitions
\begin{eqnarray}
G_{\alpha \beta}^t (x,x';t)& = & -i \langle T \Psi_{\alpha}(x,t)
\Psi_{\beta}^{\dagger}(x',0) \rangle, \\
G_{\alpha \beta}^{\tilde t} (x,x';t) & = & -i \langle \tilde T
\Psi_{\alpha}(x,t) 
\Psi_{\beta}^{\dagger}(x',0) \rangle,
\end{eqnarray}
where $T$ and $\tilde T$ are the time-ordering and anti-time-ordering
operators, respectively.
$G^>$ is the complementary
correlation function to $G^<$:
\begin{equation}
G_{\alpha\beta}^>(x,x';t)= -i \langle \Psi_{\alpha}(x,t)
\Psi_{\beta}^{\dagger}(x',0) \rangle.
\end{equation}
The corresponding matrix perturbation expansion to first order
 in the coupling $J$  yields
\begin{eqnarray}
& &{\bf G}_{(1)\alpha \beta}(x_p,x_p;t-t') = {\bf G}_{(0) \alpha
\beta}(x_p,x_p;t-t') \delta_{\alpha , \beta} \nonumber \\
& -& \frac J 2  \sum_{q=1,2} {\bf S}_q \cdot 
{\mbox {\boldmath $\sigma$}}_{\beta \alpha} \int dt_1
 {\bf G}_{(0) \alpha \alpha}(x_p,x_q;t-t_1){\bf G}_{(0) \beta \beta}
(x_q,x_p;t_1-t'). \label{i5}
\end{eqnarray}

From (\ref{i5}) the first order exchange terms of the function
 $G^<$ can be resolved as
\begin{eqnarray}
& & G^<_{\hspace{1mm}(\rm 1ex)\alpha\beta}(x_p,x_p;t-t') = - \frac J 2
{\bf S}_q \cdot {\mbox {\boldmath $\sigma$}} _{\beta \alpha} \int dt_1
\left[ G^<_{(0)}(x_p,x_q;t-t_1) G^a_{(0)}(x_p,x_q;t_1-t')
\right. \nonumber \\ 
 & + & \left. G^r_{(0)}(x_p,x_q;t-t_1) G^<_{(0)}(x_p,x_q;t_1-t')
\right] 
\hspace{5mm} (p \in \{ 1,2 \},q \not= p ) \label{i6},
\end{eqnarray}
where we have now dropped the spin indices in the unperturbed
Green's functions since they are spin independent.
To obtain (\ref{i6}), we have used appropriate relations between the
six Green's 
functions $G^<$, $G^>$, $G^t$, $G^{\tilde t}$,
$G^r$ and $G^a$ in 
order to replace the dependence of the result on $G^t$ and $G^{\tilde
t}$ by one on $G^r$ and $G^a$ (cf. Ref.\ \onlinecite{bim}). 

One relation arising from this transformation that holds
generally for operators in the Keldysh formalism and which is
particularly 
useful to us is
\begin{equation}
 (AB)^< = A^<B^a+A^rB^<, \label{i6_5}
\end{equation}
where $AB$ is to be understood as a matrix product of two general
operators $A$ and $B$, implying integrations over space and
time where applicable.
By means of
Fourier transformation of (\ref{i6}) into the frequency domain,
 (\ref{i1}) can therefore be expressed as
\begin{eqnarray}
E_{\rm RK}& = & J^2  {\bf S}_1  \cdot {\bf S}_2
\int_{- \infty}^{\infty}  {\rm Im} \left[ G^<_{(0)}(x_1,x_2)
\right. \nonumber \\ 
 & \times & \left. \{ G^r_{(0)}(x_2,x_1) + G^a_{(0)}(x_2,x_1) \}
\right] \frac {d\omega} {2 \pi}.
 \label{i7}
\end{eqnarray}
The above equation is the general expression for the RKKY interaction
out of equilibrium in a purely 1D system without translational
invariance. It is one of the central results of the
present work and the principal goal of the following treatment is to 
evaluate it explicitly for various situations.

 In an equilibrium situation it is easily
 established that (\ref{i7}) goes over to the well known
result\cite{bi7,bi8} 
\begin{eqnarray}
E_{\rm RK} & = & - \frac 1 {\pi} J^2 {\bf S}_1  \cdot {\bf S}_2
  \int_{-\infty}^{\infty}
n_F(\omega) \nonumber \\
& \times & {\rm Im} \left[ G^r_{(0)}(x_1,x_2) G^r_{(0)}(x_2,x_1)
\right] d\omega. 
 \label{i8}
\end{eqnarray}
Before we establish how the  Keldysh Green's functions $G_{(0)}^<$
and $G_{(0)}^{r/a}$ are
calculated in a nonequilibrium situation we will generalize
the results just obtained to higher dimensions. 

\subsection{Generalization to two and three dimensions}

For the extension to higher dimensions we assume that the system is
translationally invariant in the 
further one or two dimensions, corresponding to a perfectly planar
barrier.
Once we have found the solution for the purely 1D Green's function
$G^{1\rm D}_{(0)}$,
we can simplify the solution of the planar extension of the equivalent
problem by exploiting the translational invariance of the system in
the additional directions parallel to the barrier by means of
corresponding  
Fourier transforms
\begin{equation}
G^{\rm 2/3D}_{(0)}(x,x';{\bf k}_{\|};\omega)= \int d^{d-1} {\bf r}
e^{i {\bf k}_{\|} \cdot {\bf r}} G^{\rm 2/3D}_{(0)}({\bf x},{\bf
x}';\omega), \label{s5_5}  
\end{equation}
where ${\bf x}$ is a $d$-dimensional coordinate, 
${\bf r}={\bf x}_{\|}-{\bf x}'_{\|}$ is the
$(d-1)$-dimensional relative position vector parallel to the barrier
and ${\bf 
k}_{\|}$ denotes the $(d-1)$-dimensional electron 
wavevector parallel to the  
barrier, where $d\in \{2,3\}$. In the following we shall drop the
superscripts again that were just introduced to indicate the
dimensionality of the system which the corresponding Green's functions
describe, since it will be always recognizable from the arguments of
these functions whether they pertain to a purely 1D system or to a 
planar version in higher dimensions.

The extension to 2/3D leaves the spatial dependence of the Hamiltonian
unaffected in the direction perpendicular to the barrier and, as will
be shown in Sec.\ IV, only amounts to
a change in the energy arguments of the corresponding
Green's functions (see also Ref.\ \onlinecite{bi18}). 
The expression for the RKKY interaction between two arbitrarily placed
impurities with respect to the barrier in $d$-dimensions
can then be written as an extension from (\ref{i7})
\begin{eqnarray}
& E_{\rm RK} & =  J^2  {\bf S}_1  \cdot {\bf S}_2
\int \frac {d^{d-1} {\bf q}_{\|} d^{d-1} {\bf k}_{\|}} {(2
\pi)^{(2d-2)}} e^{ \displaystyle -i{\bf q}_{\|} \cdot 
{\bf r}_{12}} \nonumber\\
& \times & \int_{- \infty}^{\infty} {\rm Im} \left[
G^<_{(0)}(x_1,x_2;{\bf q}_{\|}-{\bf k}_{\|};\omega) \right.
 \nonumber \\
 & \times & \left. \{ G^r_{(0)}(x_2,x_1;{\bf k}_{\|};\omega) +
G^a_{(0)}(x_2,x_1;{\bf k}_{\|};\omega) \} \right] \frac {d\omega}
{2 \pi}
 \label{s6} \\
& &(d \in \{2,3\}), \nonumber
\end{eqnarray}
where ${\bf r}_{12}$ denotes the impurity displacement in the
direction parallel to the barrier.

The most immediate application of the present theory is in a 3D
tunneling 
system where the interaction between either two monolayers or two
slabs of spins across a barrier is of 
interest for the description of giant magnetoresistance
phenomena\cite{bi4,gm1,gm2,gm3}.  
It will be assumed that all spins within one
monolayer or slab have the same orientation due to a predominant
ferromagnetic coupling over short distances.
In equilibrium and without a barrier it was 
pointed out by Yafet\cite{bi22}
 how the interaction between two
monolayers can be obtained from the conventional 3D version. In our
case the 
corresponding expression is obtained through integrating over all 
${\bf r}_{12}$ and subsequently over ${\bf q}_{\parallel}$ in
(\ref{s6}): 
\begin{eqnarray}
& E^{\rm m}_{\rm RK}& =  \left( J \rho_{s-d}^{2D} \right)^2 \nu^{2D}  
{\bf S}_1  \cdot {\bf S}_2
\int \frac {d^2 {\bf k}_{\|}} {(2 \pi)^{2}} \int_{- \infty}^{\infty}
{\rm Im} \left[
G^<_{(0)}(x_1,x_2;-{\bf k}_{\|};\omega) \right.
 \nonumber \\
 & \times & \left. \{ G^r_{(0)}(x_2,x_1;{\bf k}_{\|};\omega) +
G^a_{(0)}(x_2,x_1;{\bf k}_{\|};\omega) \} \right] \frac {d\omega}
{2 \pi},
 \label{s7}
\end{eqnarray}
where $E^{\rm m}_{\rm RK}$ is now the monolayer interaction energy
across the  
junction, $\nu^{2D}$ is the surface area of the junction and
$\rho_{s-d}^{2D}$ 
is the surface density of the $s$-$d$ spins in each of the two planes 
parallel to the barrier. 

The situation where two finite slabs of spins
interact across a finite spacer layer which contains the barrier
is shown schematically in Fig.\ \ref{FIG2}. The interaction
between two slabs of spins across a planar tunneling barrier can be
obtained by summing the contributions coming from each pair of
interacting monolayers involved. In the present case where a
continuous system is considered this amounts to two spatial
integrations of the monolayer interaction in (\ref{s7}) in the
direction perpendicular to the barrier.
If, for
simplicity,  we consider two interacting slabs of the same thickness
$l$ which are symmetrically displaced from the barrier, these two
integrations assume the form
\begin{equation}
E^{\rm s}_{\rm RK}= \left( \rho^{3D}_{s-d} \right)^2
\int_{L-(d+l)}^{L-d} dx_1  
 \int_{R+d}^{R+(d+l)} dx_2 
\left[ \frac {E^{\rm m}_{\rm RK}(x_1,x_2)} {\left( \rho^{2D}_{s-d}
\right)^2} 
  \right], \label{s9}
\end{equation}
where $E^{\rm s}_{\rm RK}$ is the interaction energy of the two slabs
and 
in $E^{\rm m}_{\rm RK}(x_1,x_2)$, is taken from (\ref{s7}) where $x_1$
indicates  
a position within the left electrode and $x_2$ one within the right
one, respectively, and $\rho^{3D}_{s-d}$ is the 3D density of $s$-$d$
spins  
within the slabs.

Now that formal expressions for the RKKY interaction are in hand, the
next step is the evaluation of the Keldysh Green's functions 
$G_{(0)}^<(x,x')$ and $G_{(0)}^{r/a}(x,x')$ of the nonequilibrium 1D
system and their extensions to higher dimensions.

\section{Evaluation of the relevant Green's functions} 

It has been established by Caroli\cite{bi13,bi14,bi14_5,bi15} and
Feuchtwang 
\cite{bi16,bi17,bi18}, that a tunneling
system out of equilibrium can be treated by partitioning the
system into several
uncoupled parts which are considered to be in equilibrium
 at $t= -\infty$. These parts are
subsequently coupled to each other through appropriate transfer terms,
in conjunction with an adiabatic switching procedure, to finally yield
a nonequilibrium steady state. 
Formally, this corresponds to a perturbation expansion of these
transfer terms to all orders\cite{bi13,bi14},
 but it can also be shown to be  equivalent to applying Green's
theorem at the  
partitioning points\cite{bi16,bi17}.
The number of partitions to be made in
 the system is in principle
arbitrary and will largely depend on the geometry considered. In the
present case with a possibly sloping tunneling barrier of finite
width, a partitioning into three regions at the electrode-barrier
interfaces $L$ and $R$ -- shown by the vertical dashed lines in Fig.\
1 -- is most convenient. In the present treatment we will 
follow closely the approach of Ref.\ \onlinecite{bi17} for continuous
systems. The  Hamiltonian $H_0$ in (\ref{m1}) is consequently written
as a sum of three independent parts 
\begin{eqnarray}
& & H_0(x) = \Theta(L-x)H_L(x) \label{m2} \\
& + & \Theta(x-L)\Theta(R-x) H_B(x) + \Theta(x-R)H_R(x). \nonumber
\end{eqnarray}

\subsection{Green's functions in one dimension}

In one dimension the Green's function $G_{(0)}$ of the system
including the barrier has to satisfy the inhomogeneous Schr{\"o}dinger
equation 
\begin{equation}
[\omega-H_0(x)]G_{(0)}(x,x';\omega)= \delta(x-x'). \label{m2_5}
\end{equation}
Similarly one can define Green's functions for the several
uncoupled sub-parts of the system $\eta\in\{L,B,R\}$ as
\begin{equation}
[\omega-H_{\eta}(x)]g_{\eta}(x,x';\omega)= \delta(x-x'),
\label{m3}
\end{equation}
where $x$, $x'$ lie within the appropriate region determined by
 the choice of $\eta$.
Additionally, these functions have to satisfy appropriate
boundary conditions at the electrode-barrier interfaces $R$ and
$L$. In principle these boundary conditions can be an arbitrary
mixture of the 
Dirichlet and Neumann type. However, for simplicity it is best to
use one or the other exclusively, and we choose here 
the Dirichlet conditions
that the Green's functions vanish if they are taken with one of their
arguments on the respective interfaces $R$ or $L$.
Relations between these uncoupled Green's functions $g_{\eta}$ and the
Green's function $G_{(0)}$ of the 
full system, satisfying (\ref{m2_5}), can be established, as noted
before, by means of Green's theorem applied at the interfaces. If in
addition the discontinuity conditions 
for the derivatives of the full Green's function $G_{(0)}$ are used
\begin{equation}
\left. \partial_x G_{(0)}(x,x') \right\arrowvert^{x=x^{\prime
+}}_{x=x^{\prime -}}= 2m = \left. \partial_{x'} G_{(0)}(x,x')
\right\arrowvert^{x^{\prime}=x^+}_{x^{\prime}=x^-} \label{m5},
\end{equation}
(where we have left out the energy argument for brevity),
together with appropriately chosen versions of further continuity
conditions  
\begin{eqnarray}
& & \left. G_{(0)}(x,x') \right\arrowvert^{x=x^{\prime
+}}_{x=x^{\prime -}} = 0, \label{m5_5} \\
& & \left. \partial_x \partial_{x'} G_{(0)}(x,x')
\right\arrowvert^{x=x^{\prime 
+}}_{x=x^{\prime -}} = 0,  \label{m5_6}
\end{eqnarray}
which this function has to satisfy, one can find $G_{(0)}(x,x')$.
In the simple cases where $x,x' \in \{ L,R \}$,  the full Green's
function can be written as\cite{bi17} 
\begin{eqnarray}
& & \left( {\begin{array}{cc}
G_{(0)}(L,L) & G_{(0)}(L,R) \label{m6} \\
G_{(0)}(R,L) & G_{(0)}(R,R)
\end{array}} \right)
= \\
& 2m & \left( { \begin{array}{cc}
\gamma_L(L,L)+\gamma_B(L,L) & - \gamma_B(L,R) \\
- \gamma_B(R,L) & \gamma_R(R,R) + \gamma_B(R,R) 
\end{array}}\right)^{-1} = \nonumber \\
& \displaystyle{\frac {2m} D} & \left( { \begin{array}{cc}
\gamma_R(R,R)+\gamma_B(R,R) & \gamma_B(L,R) \\
\gamma_B(R,L) & \gamma_L(L,L) + \gamma_B(L,L) 
\end{array}}\right), \nonumber
\end{eqnarray}
where 
\begin{eqnarray}
D &= & [\gamma_R(R,R)+\gamma_B(R,R)][\gamma_L(L,L) +
\gamma_B(L,L)] \nonumber \\
&-&
\gamma_B(L,R)\gamma_B(R,L) \label{m7}
\end{eqnarray}
and
\begin{equation}
\gamma_{\eta}(a,b) =- \frac 1 {2m} \partial_x \partial_{x'}
g_{\eta}(x,x';\omega) \mid_{\stackrel {\scriptstyle  x=a} {x' =b}}.
\end{equation}
The above Green's function is readily cast into a retarded or advanced
version by analytically continuing $\omega \rightarrow \lim_{\delta
\rightarrow 0^+} \omega \pm i \delta$.

The calculation of
$G_{(0)}^<$, however, is significantly complicated through the matrix
property (\ref{i6_5}) inherent to all its defining equations in terms
of continuity conditions. The appropriate choice for these continuity
conditions are similar to the ones for $G_{(0)}^{r/a}$ with the only
important difference that all first derivatives are
continuous also now at $x=x'$. For a general situation with two
partitions we find that
\begin{eqnarray}
& & \left( {\begin{array}{cc}
G_{(0)}^<(L,L) & G_{(0)}^<(L,R) \label{i9} \\
G_{(0)}^<(R,L) & G_{(0)}^<(R,R)
\end{array}} \right)
= - \frac {2m} {\mid D \mid^2} \\
& \times &  \left( {\begin{array}{cc}
\gamma_R^r(R,R)+\gamma_B^r(R,R) & \gamma_B^r(L,R) \\
\gamma_B^r(R,L) & \gamma_L^r(L,L) + \gamma_B^r(L,L) 
\end{array}} \right) \nonumber \\
& \times & \left( { \begin{array}{cc}
\gamma^<_L(L,L) + \gamma^<_B(L,L)& - \gamma^<_B(L,R) \\
- \gamma^<_B(R,L) & \gamma^<_R(R,R)+ \gamma^<_B(R,R)\\
\end{array}}\right) \nonumber \\
& \times &  \left( {\begin{array}{cc}
\gamma_R^a(R,R)+\gamma_B^a(R,R) & \gamma_B^a(L,R) \\
\gamma_B^a(R,L) & \gamma_L^a(L,L) + \gamma_B^a(L,L) 
\end{array}} \right) \nonumber ,
\end{eqnarray}
where it should be noted that despite the presence of the terms
with $\gamma^<_B$ in the matrix in the middle, the full Green's
function $G_{(0)}^<$ does not depend on these terms. In a true
tunneling situation, where one only considers energies lower than the
height of the barrier, 
 this is immediately seen to be true since for those energies
no states exist in the barrier region. Another case, however,
is the one where there are states inside the barrier region in the
energy range considered, such as
impurity levels or quasi-bound states     
in a double-barrier tunneling structure; or when a
scattering problem across a non-periodic potential is considered
rather than a tunneling situation. For these cases one can
show\cite{bi14_5}
that terms in $G_{(0)}^<$ which contain $\gamma^<_B$, occur in 
conjunction with
terms coming from $\vert D \vert^{-2}$ and other terms from the matrix
product in (\ref{i9}) such
that they fulfill an identity 
$\Delta \delta( \Delta ) \Delta \equiv 0$, where $\Delta$ is the
denominator of $\gamma_B$, which is independent of the spatial
arguments of $\gamma_B$.  Put another way, the
contributions of the poles in
$\gamma_B$ which constitute $\gamma_B^<$ are suppressed in
$G_{(0)}^<$, since wherever they occur they are given zero weight.

One can understand this cancellation from a physical picture of
how the nonequilibrium system is established. 
While the left and right leads are modeled
as semi-infinite grand canonical ensembles with (possibly different)
chemical potentials on either side at $t=-\infty$, 
the barrier region is finite and not coupled to any exterior
reservoir. Once a steady state is established, 
any quantity of the fully coupled system will depend only
on the initial occupations of the semi-infinite leads.
In particular, there can be no 
dependence of $G_{(0)}^<$ on $\gamma^<_B$, since the infinite, fully
coupled system cannot remember 
the initial occupation of the finite barrier region.
In the same way it turns out that in terms containing $\gamma^{r/a}_B$
only the real part $\gamma_B$ contributes to $G_{(0)}^<$, as
contributions 
coming from the imaginary parts cancel for the reasons outlined
above. Therefore we can set $\gamma^<_B\equiv 0$ and leave away the
superscripts 
in $\gamma^{r/a}_B$ for further calculations.

For the evaluation of (\ref{i9}), however, we still need to know the
appropriate expressions for $\gamma_{\eta}^<$, $ \eta \in \{ R,L
\}$. Since the left and right decoupled
regions are separately in equilibrium, they have well defined
electron occupations $n_F^{\eta}(\omega)$, and therefore  we find that
$g_{\eta}^<$ can be expressed as 
\begin{equation}
g_{\eta}^<(x,x';\omega) = n_F^{\eta}(\omega) \left[
g_{\eta}^a(x,x';\omega) - g_{\eta}^r(x,x';\omega) \right],
\label{i9_5} 
\end{equation}
which immediately transfers to the $\gamma_{\eta}^<$ as
\begin{equation}
\gamma_{\eta}^<(x,x';\omega) = n_F^{\eta}(\omega) \left[
\gamma_{\eta}^a(x,x';\omega) - \gamma_{\eta}^r(x,x';\omega) \right].
 \label{i9_6}
\end{equation}

If the impurities lie deeper within the electrodes on the left and
right hand sides of the barrier, the relevant Green's functions
$G_{(0)}^{r/a}(x_1,x_2)$ and $G_{(0)}^<(x_1,x_2)$ can be expressed
in terms of the full Green's functions between the interfaces,
(\ref{m6}) and (\ref{i9}), and the Green's functions of the uncoupled 
leads, 
\begin{eqnarray}
 & & G_{(0)}^{r/a}(x_1,x_2) =  -(2m)^{-2} \left. \partial_{x'}
g^{r/a}_L(x_1,x') \right\arrowvert_{x'=L^-}\label{i11}\\
& \times & G_{(0)}^{r/a}(L,R) \left. \partial_{x'} g^{r/a}_R(x',x_2)
\right\arrowvert_{x'=R^+}, \nonumber
\end{eqnarray}
\begin{eqnarray}
& & G_{(0)}^<(x_1,x_2) = -(2m)^{-2} \label{i12} \\
& \times & \left\{ \partial_{x'}
g^<_L(x_1,x')\right\arrowvert_{x'=L^-}  
 G_{(0)}^a(L,R) \left. \partial_{x'} g^a_R(x',x_2)
\right\arrowvert_{x'=R^+}  \nonumber \\
& + & \left. \partial_{x'} g^r_L(x_1,x')
\right\arrowvert_{x'=L^-} G_{(0)}^<(L,R) \left.
 \partial_{x'} g^a_R(x',x_2) \right\arrowvert_{x'=R^+} \nonumber \\
& + & \left. \partial_{x'} g^r_L(x_1,x')
\right\arrowvert_{x'=L^-} G_{(0)}^r(L,R) \left. \left.
 \partial_{x'} g^<_R(x',x_2) \right\arrowvert_{x'=R^+} \right\}
\nonumber \\
& & (x_1 \leq L, \hspace{3mm} x_2 \geq R). \nonumber
\end{eqnarray}
The corresponding expressions for the Green's functions with reversed
arguments are obtained in an analogous way. 

The RKKY interaction (\ref{i7}) can now be expressed entirely in
terms of the unperturbed Green's functions of the separate subsystems
in 
equilibrium using (\ref{m6})  and (\ref{i9})-(\ref{i12}).
For the simple case where the impurities are situated
immediately on the left and right barrier-electrode interfaces
(\ref{i7}) can for example be expressed in terms of the quantities
$\gamma_\eta$ as
\begin{eqnarray}
& & E_{\rm RK} =  -2 (2m)^2 J^2 {\bf S}_1
 \cdot {\bf S}_2 \int \frac {d \omega} {2 \pi}
\vert D \vert^{-2} {\rm Re} [D^{-1}] \gamma_B(L,R) \gamma_B(R,L)
\nonumber \\ 
& \times & {\rm Im} \left\{ \left[
\gamma^r_R(R,R)+\gamma_B(R,R) \right] \gamma_L^<(L,L)
+ \gamma_R^<(R,R) \left[ \gamma_B(L,L) + \gamma^a_L(L,L) \right]
\right\}. \label{i10_5} 
\end{eqnarray}

Within the single effective mass approximation for a
barrier system as shown in Fig.\ \ref{FIG1} the functions of the
decoupled leads are found
to be
\begin{eqnarray}
& & {\displaystyle
 g^{r/a}_{L(R)}(x,x')} = +(-) \frac {2m} {q^{r/a}_{L(R)}} \label{s1}
\\ 
& \times & \left\{
{\begin{array}{cc}
\sin [q^{r/a}_{L(R)}(x - L(R))]
e^{\displaystyle \mp(\pm)iq^{r/a}_{L(R)}(x' - L(R))} & x { >(<) } x'
\\ 
\sin [q^{r/a}_{L(R)}(x' - L(R))]
e^{ \displaystyle \mp(\pm) iq^{r/a}_{L(R)}(x - L(R))} & x' { >(<) } x 
\end{array}} \right. , \nonumber
\end{eqnarray}
where $q^{r/a}_{L(R)}=\sqrt{ 2m (\omega - V^{L(R)} \pm i \delta )}$,
the upper/lower signs are associated with the superscripts r/a,
respectively, and $V^{L(R)}$ is
 the bottom of the conduction band in the corresponding side of the
junction as before.
Likewise the corresponding Green's functions for the sloping barrier
region are obtained as
\begin{eqnarray}
& & g_B(x,x') = \frac {2m \pi \kappa^{-1}} {f(R)h(L)-f(L)h(R)}
\label{s2} \\ 
& \times &  \left\{ 
{\begin{array}{cc}
\left[ h(L)f(x) - h(x)f(L) \right] \left[ f(R)h(x') - f(x')h(R)
\right] & x < x'  \\ 
\left[ h(R)f(x) - h(x)f(R) \right] \left[ f(L)h(x') - f(x')h(L)
\right] & x > x'   
\end{array}} \right. , \nonumber
\end{eqnarray}
where
\begin{eqnarray}
 f(x) & = & {\rm Ai} (\kappa x + \zeta / \eta^2),  \label{s2_7} \\
h(x) & = & {\rm Bi} (\kappa x + \zeta / \eta^2),  \nonumber 
\end{eqnarray}
are two independent solutions of the inhomogeneous Schr{\"o}dinger
equation 
\begin{equation}
\left[ \partial_x^2 - \kappa^3 x \right] f(x) =  \zeta, \label{airy}
\end{equation}
with the parameters $\kappa=-\sqrt[3]{ \frac {2meV} {R-L}}$
and $\zeta=2m\{V_0-eV/2-\omega\}$.

In equilibrium, where there is no slope to the barrier, 
one can show that $g_B(x,x')$ simplifies to
\begin{equation}
g_B(x,x')= 2m \left\{ 
{ \displaystyle \begin{array}{cc} \displaystyle
\frac { \sinh [k(x-L)] \sinh [k(x'-R)]} {k \sinh [k(R-L)]}
& x < x' \\
\displaystyle \frac { \sinh [k(x'-L)] \sinh [k(x-R)]} {k \sinh
[k(R-L)]} 
& x' > x 
\end{array}}
\right. , \label{s2_8}
\end{equation}
where $k=\sqrt{2m[V_0 - \omega]}$. Note that (\ref{s2}) and
(\ref{s2_8}) hold for all
$\omega$, i.e. including the case when $V_0 - \omega < 0$, for which
the 
sinh-functions in (\ref{s2_8}) go over to corresponding sin-functions.

When the results of the present and the previous section are
combined in equilibrium and for the limit of the barrier height or
the barrier width going to zero we can obtain analytic results for
all versions of 
the interaction considered so far. The purely 1D result from
(\ref{i7}) for this case reduces to the well known
expression for the RKKY interaction in one dimension\cite{bi22,bi2}
\begin{equation}
E_{\rm RK}=\frac {2m} {2 \pi} J^2 {\bf S}_1 \cdot {\bf S}_2 \left[
{\rm Si} (\chi) - \frac
{\pi} 2 \right] \label{s2_9},
\end{equation} 
where Si($\chi$) is the integral-sine function 
\begin{equation}
{\rm Si}(\chi)=\int_0^{\chi} \frac {\sin(\chi)} {\chi} d{\chi}
\end{equation} 
and $\chi \equiv 2 k_F x$ (with $x \equiv \vert x_2 -x_1 \vert$)
gives the phase of the characteristic oscillation of the interaction
at twice the Fermi wavevector $k_F$.

\subsection{Green's functions in higher dimensions}

Once we have found the solutions for the purely 1D Green's functions
$g_{\eta}$, which satisfy (\ref{m3}),
we can find the solution of the planar extension of the equivalent
problem, as demonstrated by (\ref{s5_5}), by means of Fourier
transforms in the directions parallel to  the barrier
\begin{equation}
[(\omega - k^2_{\|}/2m)-H_{\eta}(x)]g_{\eta}(x,x';{\bf
k}_{\|};\omega)= \delta(x-x').\label{s55} 
\end{equation}
Here again retarded and advanced versions of $g^{\rm
2/3D}_{\eta}(x,x';{\bf k}_{\|};\omega)$ can be found simply through
continuing analytically
\begin{equation}
g^{r/a}_{\eta}(x,x';{\bf k}_{\|};\omega)= g_{\eta}(x,x';{\bf
k}_{\|};\omega \pm i \delta). \label{m3_5} 
\end{equation}
Correspondingly we find
\begin{equation}
g^<_{\eta}(x,x';{\bf k}_{\|};\omega) =
n_F^{\eta}(\omega) \left[ g^a_{\eta}(x,x';{\bf
k}_{\|};\omega) - g^r_{\eta}(x,x';{\bf k}_{\|};\omega)
\right], \hspace{5mm} \eta \in \{L,R\}. \label{m3_6}
\end{equation}
It is important to realize at this point that for the retarded and
advanced Green's functions $g^{r/a}_{\eta}$ the extension to higher
dimensions 
only leads to a shift in the energy argument $\omega \rightarrow
\omega - k_{\|}^2/2m$, as can be seen from the form of their defining
equation (\ref{s55}), i.e.
\begin{equation}
g^{r/a}_{\eta}(x,x';{\bf k}_{\|};\omega) = g^{r/a}_{\eta}(x,x';\omega
-  k_{\|}^2/2m ). \label{s5_7} 
\end{equation}
This property is seen to also translate in part to the functions
$g^<_{\eta}$ with the only important difference that the
energy arguments of the occupation functions remain unchanged.

By using the properties (\ref{m3_6}) and (\ref{s5_7}) one can
establish that the Green's functions of the full system
$G^{r/a}_{(0)}$ and $G^<_{(0)}$ can be represented in the following
way: 
\begin{eqnarray}
G^{r/a}_{(0)}(x,x';{\bf k}_{\|};\omega) & = &
G^{r/a}_{(0)}(x,x';\omega -  k_{\|}^2/2m ), \label{s57} \\ 
G^<_{(0)}(x,x';{\bf k}_{\|};\omega) & = & n^L_F(\omega)
\Gamma_L (x,x';\omega -  k_{\|}^2/2m ) + n^R_F(\omega) \Gamma_R
(x,x';\omega - k_{\|}^2/2m ) \label{s5_8}, 
\end{eqnarray} 
where the functions $\Gamma_{\eta}$,  $\eta \in \{ L, R \}$
associated with the Fermi functions $n^{\eta}_F$ can be written
 in terms of sums and products of spatial
derivatives of the $g^{r/a}_{\eta}$, $\eta \in \{ R, B, L
\}$ when (\ref{m6})-(\ref{i12}) are used to divide $G^<_{(0)}(x,x')$
into contributions containing left or right occupation functions only.
For example one finds:
\begin{eqnarray}
\Gamma_{L(R)}(L,R) & = & - \frac {2m} {\vert D \vert^2} \gamma_B(L,R)
\left\{ 
\gamma^{r(a)}_{R(L)}[R(L),R(L)]+ \gamma_B[R(L),R(L)] \right\}
\nonumber \\ 
& \times & \left\{ \gamma_{L(R)}^a[L(R),L(R)] -
\gamma_{L(R)}^r[L(R),L(R)] \right\}. \label{s5_9}
\end{eqnarray}

The properties (\ref{s57}) and (\ref{s5_8}) prove to be useful for the
evaluation of the frequency and wavevector integrations in the
expression for the monolayer interaction (\ref{s7}), which is also
needed for the calculation of the slab interaction in (\ref{s9}).
If the interaction is considered at zero temperature and the
additional property is used that the density of states factor coming
from the 2D ${\bf k_{\|}}$-integration in (\ref{s7}) is just a
constant, one can reduce the three integrations in  (\ref{s7}) to a
sum of two single integrations as
\begin{eqnarray}
& & \int \frac {d^2 {\bf k}_{\|}} {(2 \pi)^{2}} \int_{-
\infty}^{\infty} 
\frac {d\omega} {2\pi} \left[ \Theta (\mu - \omega) \tilde \Gamma_L
(\omega - \frac 
{k_{\|}^2} {2m} )  + \Theta(\mu - \omega - eV) \tilde \Gamma_R (\omega
- \frac {k_{\|}^2} {2m} ) \right] \nonumber \\ 
& = &  \frac {2m} {8 \pi^2} \int_{0}^{\mu} dz
(\mu-z) \tilde \Gamma_L (z)+ \int_{-eV}^{\mu-eV} dz (\mu-z-eV) \tilde
\Gamma_R (z), \label{n3}
\end{eqnarray}
where the $\Theta$-functions derive from the sharp Fermi
distributions $n_F^{\eta}(\omega)$ at $T=0$ and
\begin{equation}
\tilde \Gamma_{\eta} (z) = 
{\rm Im} \left\{ \Gamma_{\eta} (x_1,x_2; z) \left[ G^r (x_2,x_1; z)+
G^a (x_2,x_1; z) \right] \right\}.
\end{equation}
 The lower limits at $0$ and $-eV$ in (\ref{n3}) follow
from the fact that there are no states below the bottoms of the
conduction 
bands in the left and right lead, respectively.
As a result of this simplification the
interaction energy between monolayers of spins in 3D is not much
harder to evaluate than the interaction energy in the purely 1D case.

For a 3D planar junction in equilibrium without a barrier we find from
(\ref{s7}) that the interaction density for monolayers is
\begin{equation}
E^{\rm m}_{\rm RK}=\frac {2mk_F^2} {2(2 \pi)^2} 
\left( J \rho_{s-d}^{2D} \right)^2 \nu^{2D} {\bf S}_1 \cdot
{\bf S}_2  \left[ {\rm 
Si} (\chi) - \frac {\pi} 2 + 
\frac {\chi \cos (\chi) - \sin (\chi)} {\chi^2} \right],  \label{s8}
\end{equation}
which is evidently similar to the purely 1D result. 
The monolayer interaction is therefore often called quasi-one
dimensional\cite{bi22}. The slab interaction from (\ref{s9}) in this
case assumes the form\cite{bi9}  
\begin{equation}
E^{\rm s}_{\rm RK} = \frac {2m \left(J \rho^{3D}_{s-d}
 \right)^2 \nu^{2D}} {16 (2 \pi)^2} 
{\bf S}_1 \cdot {\bf S}_2 \left\{ {\cal F}(s+2l)-2{\cal F}(s+l)+{\cal
F}(s) \right\}, \label{s10} 
\end{equation}
where $\rho^{\rm 3D}_{s-d}=\rho^{\rm 1D}_{s-d}\rho^{\rm 2D}_{s-d}$
 is the 3D density of $s$-$d$ spins within the slabs,
$s=2d+R-L$ is the spacing between
the slabs, $l$ is the width of each
slab (see Fig.\ \ref{FIG2}) and ${\cal F}(x)$ is the range function
\begin{equation}
{\cal F}(x) = \chi^2 \left[ {\rm Si} (\chi) - \frac \pi 2 \right] 
+ \chi \cos (\chi) + \sin (\chi) + 2 {\rm Si} (\chi), \hspace{5mm}
\chi=2k_Fx. 
\end{equation}
Another special case worth noting in this context is the
interaction between two 
magnetic half-spaces separated by a non-magnetic spacer. By letting $l
\rightarrow \infty$ we find that 
(\ref{s10}) simplifies to 
\begin{equation}
E^{{\rm s} \infty}_{\rm RK} =  \frac {2m \left(J \rho^{3D}_{s-d}
 \right)^2 \nu^{2D}} {16 (2 \pi)^2} 
{\bf S}_1 \cdot {\bf S}_2 \left\{ \left[ \chi_s^2 + 2 \right] \left[
{\rm Si} 
 (\chi_s) - \frac \pi 2 \right] + \chi_s \cos (\chi_s) + 
\sin (\chi_s) \right\},  \label{s11}
\end{equation}
where $\chi_s=2k_Fs$.

In order to obtain solutions for (\ref{i7}), (\ref{s7}) and (\ref{s9})
also for more general cases we have to perform the corresponding
energy integrals numerically.

\section{Numerical results}

\subsection{Comparative results in equilibrium}

Numerical results for the RKKY interaction across a
tunneling junction in equilibrium have been given by Mukasa {\it et
al.\ }\cite{bi21}
 These authors
 were particularly interested in providing a theoretical model
for possible applications in exchange force microscopy, as
mentioned in the introduction, which is based on the
RKKY interaction as the dominant force between a tunneling tip and a
sample. For this purpose Mukasa {\it et al}.\ performed an approximate
version of scattering wave perturbation theory for free electrons in a
1D system
by including the transmission coefficient of an electron tunneling
through 
the barrier at an intermediate stage in their calculation. 
In contrast,
our analytic expression for the RKKY interaction is exact          
(to order $J^2$) since the constituent Green's functions include 
scattering by the barrier exactly. 

In order to compare our results in equilibrium with the ones in 
Ref.\ \onlinecite{bi21}, we have implemented our calculation with the
same model parameters using a Fermi energy of 
$\mu \equiv \mu^L=\mu^R=5.0 {\rm eV}$ in
equilibrium and a lattice constant $a_0=2.50 {\rm \AA}$, along with a
Fermi wavevector $k_F=1.26 \times 10^{10} 
 {\rm m}^{-1}$, implying a relative effective electron mass of
$m/m_{\rm e}=1.20$, where $m_{\rm e}$ is the bare electron mass. Fig.\
\ref{FIG3}(a) shows plots representing the dimensionless 
interaction range function $\Phi(x)$ with
\begin{equation}
\Phi(x)=- \pi \left\{ \frac {2m} {2\pi} J^2 {\bf S}_1
\cdot {\bf S}_2 \right\}^{-1} E_{\rm RK}(x), \label{n2}
\end{equation}
where $ E_{\rm RK}(x)$ was obtained from (\ref{i7}) as evaluated in 
(\ref{i10_5}).  
 In Fig.\ \ref{FIG3}(a), the
impurities are 
considered to be fixed at the electrode-barrier interfaces, 
while the width
of the barrier is continuously increased from $0.0{\rm \AA}$ to
$5.0{\rm \AA}$, to
represent for example the height of an STM tip above a sample.
Without the barrier ($V_0/\mu=0.0$), the range function reduces to
$\Phi(x)= \pi [\pi/2 - {\rm Si}(\chi)]$, where $\chi= 2 k_Fx$.

We have also plotted in Fig.\ \ref{FIG3}(a) the interaction for 
this case and for a barrier height of $V_0/\mu= 0.5$, 
where the system is
in a scattering state.
We find that as the barrier height is increased from zero,
the interaction varies with a longer
wavelength, corresponding to a decreasing relative wavevector between
the top of the barrier and the Fermi level $ K_F =
\sqrt{2m(\mu-V_0)}$.

 For $V_0/\mu > 1$,  the strength of the
interaction decays  exponentially with the width of the barrier, 
with an
exponent that increases with $V_0/\mu$, and the crossover into the 
antiferromagnetic (AFM) regime is lost once $V_0/\mu \geq 1.3 $.
In the interval of $V_0/\mu \in
[1.0,1.3]$, the interaction still experiences one slight crossover to
the AFM regime. This can be understood to arise from the
nature of the transmission and reflection coefficients of the barrier,
which are not purely exponential, but a mixture of hyperbolic
functions.

However, our results explicitly {\it do not} show the large
oscillation 
of the interaction far in the ferromagnetic (FM) regime as was
obtained in Ref.\ \onlinecite{bi21}, markedly for their curve 
$V_0/\mu = 1.05 $. Such a 
behavior is unphysical in a genuine tunneling situation, 
and reflects the approximate treatment of the
transmission of scattered waves through the barrier in 
Ref.\ \onlinecite{bi21}.
Our Green's function approach in comparison includes
the single particle barrier potential fully from the beginning and
therefore allows an exact evaluation of the RKKY interaction.

Other interesting cases to investigate are the interaction of two
monolayers, or of two semi-infinite slabs of 
magnetic impurities in the presence of a
tunneling barrier.
Such systems could for example be realized by
coating both sides of a fixed or mobile tunneling junction 
with a magnetic material. 
For these cases one can make use of (\ref{s7}) for monolayers and
of (\ref{s7}) and (\ref{s9}) for slabs, both in conjunction with
(\ref{n3}). We shall consider (\ref{s7})
in equilibrium and  set $eV=0$ in (\ref{n3}). In Fig.\
\ref{FIG3}(b) and Fig.\ \ref{FIG3}(c) we show results for the
interaction range functions
\begin{eqnarray}
\Phi_{\rm m}(x)& = & - \pi \left\{ \frac {2mk_F^2} {2(2 \pi)^2} 
\left( J \rho_{s-d}^{2D} \right)^2 \nu^{2D} {\bf
S}_1 \cdot {\bf S}_2 \right\}^{-1}
E^{\rm m}_{\rm RK}, \label{n4} \\
\Phi^{\infty}_{\rm s}(x) & = & - \frac \pi 2
 \left\{\frac {2m \left(J \rho^{3D}_{s-d}
 \right)^2 \nu^{2D}} {16 (2 \pi)^2} 
{\bf S}_1 \cdot {\bf S}_2 \right\}^{-1} E^{{\rm s} \infty}_{\rm RK},
\end{eqnarray}
 for monolayers and for slabs, 
respectively, which were obtained for a planar version of the
arrangement used 
for Fig.\ \ref{FIG3}(a). From (\ref{s8}), for zero barrier height,
$\Phi_{\rm m}$ reduces to $\Phi_{\rm m}= \pi
\{ \pi/2 - {\rm Si} (\chi) - [ \chi \cos (\chi) - \sin (\chi) ]/\chi^2
\}$ and $\Phi^{\infty}_{\rm s}$ to $\Phi^{\infty}_{\rm s}=- \pi/2 \{ [
\chi^2 + 2][ {\rm Si} (\chi) - \pi /2 ] + \chi \cos (\chi) + 
\sin (\chi) \}$. 
The curves in Fig.\ \ref{FIG3}(b) show that characteristically
 the monolayer interaction $\Phi_{\rm m}$ decays much faster than the
purely 1D interaction $\Phi$, shown in Fig.\ \ref{FIG3}(a), 
when compared
with equal parameter values. Especially for $V_0/\mu\geq1.0$,
$\Phi_{\rm m}$ shows a much stronger decay than $\Phi$. This can be
understood as a result of the $k_{\|}$-integration in (\ref{s7}) which
effectively averages the interaction of one spin on one side of the
junction with all other spins on the opposite side. This average
which extends over many oscillatory contributions leads to destructive
interference effects which results in the observed damping
of $\Phi_{\rm m}$ relative to $\Phi$.

When the RKKY interaction of infinite half-spaces represented by
$\Phi^{\infty}_{\rm s}$ is compared to $\Phi_{\rm m}$ and $\Phi$ it is
seen that the double spatial integration from (\ref{s9}) increases the
relative oscillation strength of $\Phi^{\infty}_{\rm s}$ when compared
to $\Phi_{\rm m}$, and at $V_0/\mu=1.05$ the relative cross-over
$\Phi_{\rm s}$ into the AFM regime becomes even stronger than the one
for $\Phi$.

\subsection{Nonequilibrium behavior}

\subsubsection{Impurities on the electrode-barrier interfaces}

In the following we establish how the results shown in Fig.\
\ref{FIG3} for the interaction in equilibrium are 
modified when a finite
bias is applied to the junction.
Fig.\ \ref{FIG4}(a) and Fig.\ \ref{FIG4}(b) show the 
1D interaction between
two magnetic impurities  placed on opposite 
electrode-barrier interfaces for various strengths of the bias in 
conjunction with initial equilibrium ratios $V_0/\mu=1.05$ in Fig.\
\ref{FIG4}(a) and $V_0/\mu=1.5$ in Fig.\ \ref{FIG4}(b), and with
otherwise the same model parameters as in Fig.\ \ref{FIG3}(a).

As the bias is increased in Fig.\
\ref{FIG4}(a) the
interaction starts to exhibit oscillations when the right edge of the
barrier potential $V(R)=V_0-eV$ is pulled below the chemical potential
of the left hand side $\mu^L$. This oscillation arises because the
wavefunctions of high 
energy electrons tunneling from the left exit the barrier
through its sloping part, and become oscillatory over the
distance where they are above the barrier. In this regime we find that
the wavevector of the oscillation can be roughly approximated by the
wavevector of electrons tunneling from the left Fermi level at the
position of the right interface with the barrier, i.e.\
\begin{equation}
q_F = \sqrt{2m[\mu-V_0+eV]}, 
\label{n1}
\end{equation}
giving an oscillatory wavelength $\Delta x = 2 \pi / q_F$.
The wavelength of the interaction becomes smaller, i.e.\ the
value of $q_F$ in (\ref{n1}) increases, as the bias is increased. In 
Fig.\ \ref{FIG4}(a), one can see that as the bias is turned up,  the 
antiferromagnetic region initially vanishes and then almost reappears 
at a smaller distance between the impurities.

In Fig.\ \ref{FIG4}(b), for $eV/\mu=1.50$, a similar behavior 
to the one in
Fig.\ \ref{FIG4}(a), for $eV/\mu=1.05$,
 is observed. The main difference to Fig.\
\ref{FIG4}(a) is that for
$eV/\mu=1.50$ and in equilibrium the
interaction just decays exponentially with
no AFM region, whereas out of equilibrium
such an AFM region is established as
the bias is turned up. The fact that  the
oscillations caused by high bias are not centered
around the $\Phi=0$ line in Fig.\ \ref{FIG4}(a) and
Fig.\ \ref{FIG4}(b) can be understood to arise from the asymmetric
shape of the sloping barrier.
From  these figures it is apparent that
in an arrangement where the impurities are attached to the
electrode-barrier interfaces, such as an STM tip and sample, both
coated with magnetic materials,
the interaction becomes tunable between FM and AFM
by varying the bias alone. 

Since, as mentioned before, an interesting application of the present
system would be in exchange force microscopy --- 
where the force caused
by the exchange interaction on a tunneling tip in an STM is
measured --- we have plotted in Fig.\ \ref{FIG4}(c) and Fig.\
\ref{FIG4}(d) the spatial derivative $-d\Phi(x)/d x$ of the
range function $\Phi(x)$ from Fig.\ \ref{FIG4}(a) and Fig.\
\ref{FIG4}(b), respectively. Both Fig.\ \ref{FIG4}(c) and Fig.\
\ref{FIG4}(d) show explicitly that the onset of oscillations in the
interaction for an appropriate bias will lead to a force which is
alternating in sign.
An estimate of the absolute strength of the exchange force and
predictions that it should be measurable with a state of the art
 STM are postponed to a discussion in Sec.\ VI.

We next consider the effect of a finite bias on a system of
two magnetic monolayers interacting
across a 3D planar barrier.
In Fig.\ \ref{FIG4}(e) and Fig.\ \ref{FIG4}(f) we plot the range
function $\Phi_{\rm m}$, 
using (\ref{s7}), (\ref{n3}), and (\ref{n4}), for a planar version
of the arrangement used in 
Fig.\ \ref{FIG4}(a) and Fig.\ \ref{FIG4}(b).
As in Fig.\ \ref{FIG4}(a) and Fig.\ \ref{FIG4}(b) the interaction
starts to exhibit an oscillatory behavior once the slope of the
barrier gets steep enough.
In both  Fig.\ \ref{FIG4}(e) and Fig.\ \ref{FIG4}(f) it is evident 
that the oscillations can reach into the AFM
region, although,  as
in Fig.\ \ref{FIG4}(a) and Fig.\ \ref{FIG4}(b), the oscillations
 are again not centered 
about the $\Phi_{\rm m}=0$ line, 
but are shifted into the FM region.

One more remark is in
order when nonequilibrium results for monolayers are compared to
results in equilibrium. In equilibrium the total interaction energy
$E^{\rm m}_{\rm RK}$ in (\ref{s7}) is proportional to $\mu$, and we
have normalized the range function $\Phi_{\rm m}$ in
Fig.\ \ref{FIG4}(e) and  Fig.\ \ref{FIG4}(f)
with respect to the equilibrium Fermi energy
$\mu$. However, the total nonequilibrium interaction from (\ref{s7})
depends in magnitude on a non-separable mixture 
of $\mu$, $eV$, $\mu+eV$ and 
$\mu-eV$. This would be evident if we had plotted 
the interaction back to
$x=0$ where the results for various strengths of 
$eV$ would no longer converge
in a single point, as they do in equilibrium for different
$V_0/\mu$. 

Altogether,  our results in  Fig.\ \ref{FIG4}(a) and  Fig.\
\ref{FIG4}(b) as well as in Fig.\ \ref{FIG4}(e) and 
Fig.\ \ref{FIG4}(f)
show that when the impurities are attached to the
electrode-barrier interfaces a switching of the interaction can be
achieved in many situations by changing the bias alone, but in general
this switching behavior depends quite
strongly on the particular properties of the barrier.
In the following we will show that such a switching is much more
reliably achieved by placing the single spins or layers of
spins within the electrodes a finite
distance away from the interfaces with the barrier. 

\subsubsection{Impurities within the electrodes}

We now consider the nonequilibrium behavior of the interaction 
when the impurities are placed inside the electrodes a distance $d$
away from the electrode-barrier interfaces. 
In Fig.\ \ref{FIG5}(a) we show a
surface plot of a 1D arrangement with the relative barrier height
$V_0/\mu=1.5$, where the barrier width is 
kept constant at $1.0 {\rm \AA}$
[cf. corresponding point in Fig.\ \ref{FIG4}(b)]. The impurities are
now moved away from their initial 
positions on the electrode-barrier interfaces to a maximum
distance of $d=5.0 {\rm \AA}$ from either 
interface (plotted across the
figure).
At the same time the bias is increased from $eV/\mu=0.0$ to
$eV/\mu=1.5$ (plotted into the depth). We have overlayed a contour
plot to make it easier to 
identify which regions of the surface lie in the FM or AFM
regime. 
The solid zero contour line indicates the boundaries of these
regions. In the same way,  we show in Fig.\ \ref{FIG5}(b)
the interaction density between two monolayers and in Fig.\
\ref{FIG5}(c) the interaction density between two slabs of finite
thickness which are placed a finite distance inside two 3D electrodes.
 The range function in Fig.\ \ref{FIG5}(b)
is again normalized with respect to the equilibrium Fermi
energy $\mu$. The case of interacting slabs turns out to have
quite special features which will be discussed later in this section.

One can see that in both Fig.\ \ref{FIG5}(a) and 
Fig.\ \ref{FIG5}(b) the
interaction is already oscillatory in 
equilibrium as $d$ is varied, even though 
$V_0/\mu \geq 1.0$, since now
the electrons
have to travel over a finite region on either side
of the junction where their wavefunctions are oscillatory. 
The presence of the
barrier in this case  leads
to an overall exponential damping which reduces the strength of the
oscillations everywhere.
As the bias is turned up these oscillations evolve into an
interference between up to five contributing components which can be
 explained as follows: To the order of perturbation theory considered,
the spins on the left side of the junction interact with
 the ones on the right through electrons tunneling between 
the locations
of the spins. Electrons in the vicinity of
the left spins which are able to
perform this process are available up to the Fermi level on the
left. The wavevector $k_{LL}$ of the spin
 polarization of the conduction electrons in the left lead
is determined by the cut-offs
of the frequency integration in (\ref{i7}) and (\ref{s7}) at the Fermi
energy and at the band bottom on the left side $V^L$, so that
$k_{LL}=\sqrt{2m(\mu^L-V^L)}$. This wavevector is, however, a
different one, namely $k_{LR}=\sqrt{2m(\mu^L-V^R)}$ once the tunneling
electrons  have penetrated the 
barrier and interact with the spin on the right side.
The same process applies to the spins on the left 
feeling the presence of the
ones on the right. The corresponding wavevectors involved
 therefore comprise 
$k_{ij} = \sqrt{2m(\mu^i-V^j)},$ $ (i,j \in \{L,R\})$,
i.e. four different ones in principle. In addition to this, one also
observes a quite strong oscillation which contains the wavevector
$k_{eV}=\sqrt{2meV}$. This oscillation can be 
understood to arise from the
interval in the frequency
integration in the expression for the interaction (\ref{i7})
which extends from the band bottom of the right lead 
to the band bottom of
the left lead, i.e. over $V^L-V^R=eV$. Since all cases 
of the interaction
studied here have a one-dimensional or quasi-one-dimensional behavior,
both the densities of states of the electrons in the left lead and 
in the right lead exhibit a singular behavior at
the respective band bottoms. The existence of the singular parts
of these densities of states at the integration limits 
gives rise to the
$\sqrt{2meV}$-oscillation, which in some situations becomes 
the principal
oscillation in the nonequilibrium system.

 Assuming that $\mu^R$ moves by the same
amount $eV$ as $V^R$, and with $V^L$ normalized to zero as
shown in Fig.\ \ref{FIG1}, this set
of wavevectors reduces to a total of four,
namely $k \in \left\{ \sqrt{2m\mu^L}, 
\sqrt{2m(\mu^L+eV)}, \sqrt{2m(\mu^L-eV)} \right\}$ 
due to the sharp Fermi distributions $n^L$ and $n^R$,
and the $k_{eV}=\sqrt{2meV}$ oscillation as discussed before. 
In both Fig.\ \ref{FIG5}(a) and Fig.\ \ref{FIG5}(b), a superposition
of these principal wavevectors can be found. Once
the bias $eV$ moves the Fermi energy on the right below the
band bottom on the left, $\mu^R \leq 0$, the corresponding
components of the 
oscillations turn into an exponential decay, leaving only three
contributions to the 
oscillations.
 This transition can be seen as a kink in the
contour plots in Fig.\ \ref{FIG5}(a) and 
Fig.\ \ref{FIG5}(b) at a bias
of $eV=1.0$.
At higher biases, for arrangements where the
impurities lie very close to the barrier, i.e.\ 
when $d/(R-L) \ll 1.0$,
barrier effects such as those displayed in the 
plots of Fig.\ \ref{FIG4}(b)
and Fig.\ \ref{FIG4}(f) become more important. However, for
ratios $d/(R-L) > 1.0$ we expect these effects 
to be minimal when the width
of the barrier is held constant.
Both Fig.\ \ref{FIG5}(a) and Fig.\ \ref{FIG5}(b) show
explicitly how the interaction between two single spins in 1D and
between two layers of spins, 
respectively, is tunable between ferromagnetic and
antiferromagnetic coupling at a fixed impurity configuration
by varying the bias alone. 
Furthermore it is clear in Fig.\ \ref{FIG5}(a) and especially
in Fig.\ \ref{FIG5}(b) that the interaction falls off significantly 
less rapidly in the nonequilibrium regime. This circumstance has
particularly strong consequences for the interaction 
between slabs of spins.

In Fig.\ \ref{FIG5}(c) we have plotted the interaction 
between two finite
magnetic slabs of spins with a width of the slabs of $l=10{\rm
\AA}$ using otherwise the same conditions as were taken for the
monolayer-case shown in Fig.\ \ref{FIG5}(b). 
While it can be seen in Fig.\ \ref{FIG3}(c) that in equilibrium the
interaction for slabs converges to a finite value in the limit when
the thickness of the slabs goes to infinity, $l \rightarrow \infty$,
this is no longer the case out of equilibrium. Rather the interaction
is seen to become roughly proportional to the thickness 
of the slabs for
thicknesses  $l \gg \pi/k_F$. This can be attributed to the longer
range of the monolayer interaction as mentioned before, which
destroys the quite sensitive convergence of the double integral in
(\ref{s9}) for $l \rightarrow \infty$. 

When the slab interaction (\ref{s9}) is calculated one can
exchange the spatial integrations with the frequency and wavevector
integrations. This 
has the advantage that the spatial integration can be performed
analytically first. Since the explicit representation of the integrand
is quite involved, we have postponed it to Appendix A.
From (\ref{ap8}) it can be seen that the integrand contains several
terms which 
are proportional to $l$. In equilibrium, as shown in (\ref{ap6}),
these terms cancel one 
another, but out of equilibrium, when the frequency integration for
these terms is cut off at different points through different Fermi
occupation functions, this
cancellation ceases to be complete. Therefore, when the bias is
switched on, a residual $l$-dependence remains, which increases as the
bias is increased. The total interaction energy for slabs therefore
 scales extensively with the width of the slabs. It should be
noted that this extensive dependence of the slab interaction 
energy out
of equilibrium is not caused
by the offset of the band bottoms in our model. If an
equilibrium system is considered in which materials with different
band structures form a junction, the slab interaction continues to
converge always [cf. (\ref{ap6})].

For this reason we have
compressed the scale of the plot in Fig.\ \ref{FIG5}(c) in the
direction of the bias by a 
factor of $[1+ 40 \times eV/\mu]^{-1}$, so that, as shown, 
the height of the
oscillations stays approximately the same.
Moreover, one can see that the $k_Fd$ oscillations observed in 
equilibrium vanish completely at a very small bias and
turn into oscillations with phase  $(d/2)\sqrt{2meV}$. That this is
the case can be seen from the overlayed contour lines, which,
particularly in the upper right corner, show a
dependence $\propto 1/\sqrt{2meV}$ when taken as a function of $eV$.
Mathematically, the dominance of these oscillations can be
understood to arise from the spatial integrations in formula
(\ref{s9}). When the spatial integrations are interchanged
with the frequency integration in (\ref{s9}), as it is 
done in (\ref{ap0})
in Appendix A, the space integrations are 
effectively taken over products of the 
oscillatory, i.e.\ trigonometric, wavefunction-like expressions
$\eta^{r/a}_{L(R)}$ from (\ref{ap1}) which relate to electrons in the 
left and right leads [cf.\ (\ref{ap3})-(\ref{ap5})]. This yields an 
extra factor of $(q^{r/a}_{L(R)})^{-1}$
in $\xi^{r/a}_{L(R)}$ from (\ref{ap4}). A factor of $\xi^{r/a}_{L(R)}$
 occurs at least once in every term in the integrand of the subsequent
 frequency integration in (\ref{ap8}),
which enhances the peak in the quasi-1D density of states of 
the monolayer
system occurring on the bottom of the band in the corresponding lead.
 The frequency integration itself
very much acts as a Fourier transform into real space, 
which gives the 
peaks the effect of frequency components of the real-space
 oscillations of the interaction. These peaks are
just an energy interval $eV$ apart, which explains the occurrence of
the $k_{eV}=\sqrt{2meV}$ wavevector as a difference wavevector
 between these components. The strength of $k_{eV}$ is then determined
by the strength of the peaks, i.e.\ the amplitude of the Fourier
components.

Since, as is shown in Fig.\ \ref{FIG5}(c), the interaction between
magnetic slabs oscillates with almost only the $k_{eV}$
contribution present, it should be
controllable in a simple fashion by tuning the bias. 

\section{Discussion}

In this section, the possibilities for the experimental study of the
RKKY 
interaction in tunneling systems out of equilibrium are
discussed. As mentioned in the introduction, one possible way to
probe the behavior of the interaction that we predict
is to use an STM to measure the exchange force, such
as shown in Fig.\ \ref{FIG4}(c)
and Fig.\ \ref{FIG4}(d).

In order to estimate the absolute value of the 
exchange force, we must
estimate $J$ in the prefactor of the force function 
between two magnetic impurities, adsorbed to the STM tip and 
the sample, respectively. One method to estimate $J$, which is also
used in Ref.\ \onlinecite{bi21}, is to use the equation 
for the Kondo temperature $T_K$
\begin{equation}
T_K = \frac {\mu} {k_B} 
\exp \left[ - \frac 1 {J g_{s-d}^{\rm 1D}(\mu)} \right],
\end{equation}
where $k_B$ is the Boltzmann constant and
 $ g_{s-d}^{\rm 1D}(\mu) = (2\pi)^{-1} \sqrt{2m/\mu}$ is the 1D 
density of $s$-$d$ spin states at the equilibrium Fermi energy $\mu$.
 For measured
$T_K$'s of about
$T_K \sim 100 K $ one obtains $J g_{s-d}^{\rm 1D}(\mu) \sim
0.2$. From this the function $F(x)$ representing the total measured
exchange force is determined as
\begin{equation}
F= \frac {2m} {2 \pi^2} J^2 {\bf S}_1 \cdot {\bf S}_2
\frac {d \Phi(x)} {dx}.
\end{equation}  
From the parameters introduced above, we find
\begin{equation}
F \sim  {\bf S}_1
 \cdot {\bf S}_2 \frac {d \Phi(x)} {dx} 
\times {\cal O}\left( 10^{-20} {\rm N m} \right),
 \end{equation} 
which for the range functions plotted in Fig.\ \ref{FIG4}(c) and 
Fig.\ \ref{FIG4}(d) leads to
forces of about $F \sim {\cal O}( 10^{-11} {\rm N})$. As noted
in the literature\cite{bi21,bi41} the resolution of current atomic
force microscopy 
is about $10^{-11}{\rm N}$---$10^{-13}{\rm N}$, which means that the
exchange force should be experimentally observable.

For the observation of the RKKY interaction in layered tunneling
systems which exhibit giant magnetoresistance phenomena the results
shown in Fig.\ \ref{FIG5} should give an appropriate
prediction. Especially for the case of 3D systems the results for the
interaction between interacting slabs of impurities shown in Fig.\
\ref{FIG5}(c) 
would be most applicable. The most surprising result in this situation
is 
 that we find that the interaction scales extensively with the
thickness of the  
slabs. In principle the expression for the interaction we study 
would diverge in the limit of semi-infinite slabs,
 but of course this would not be
 observable in real systems, since processes like spin relaxation or
the  
scattering of electrons on non-magnetic impurities would eventually
impose a  
maximum range on the interaction. However, for 
slabs of reasonable thickness the switching of the bias should produce
an 
observable change in the decay behavior of the interaction, i.e.\ in a
system 
with fixed spin positions an anomalous increase of the 
interaction strength should be measurable.

Furthermore, the tunability of the RKKY interaction in a tunneling
system out of equilibrium is particularly
interesting in view  of the fact that the impurity configuration is
normally pre-set in a solid system, which means it is usually not
possible 
to directly observe the oscillatory dependence of the interaction
on the impurity spacing. With the present arrangements it becomes
possible to observe oscillations of the RKKY coupling via the
variation of wavevectors on either side of the junction as
the bias is varied. If one follows for example the direction of the
bias at a given distance of the spins in any of the contour plots
shown in Fig.\ \ref{FIG5}, one can almost always observe at least one
crossing from FM to AFM coupling or {\it vice versa}.

Since, out of equilibrium, as indicated before,
the interaction depends on several contributing wavevectors,
the relative dependence on these could for example be
investigated experimentally by using different materials for the
connecting leads and by 
placing the spins asymmetrically around the barrier, e.g.\ on the
electrode-barrier interface on one side of the junction and further
within the electrode on the other side.
As the arrangements in Fig.\ \ref{FIG5}(b) and Fig.\ \ref{FIG5}(c)
describe routinely achievable physical systems, this suggests that the
phenomena found theoretically for such systems 
should also be accessible to experiment.

\section{Conclusion}

This paper presents for the first time a theoretical treatment of the
RKKY interaction in systems out of equilibrium. Except for processes
involving 
ultrashort time dependent excitations, a proper nonequilibrium
situation seems
only routinely attainable in structured systems which include a
potential 
barrier. The main achievement of this work is that we have
obtained a proper field theoretic description of such a system which
is much more systematic than conventional scattering wave
perturbation approaches to the problem (cf.\ Ref.\ \onlinecite{bi21}).
Difficulties such as taking
proper account of the different occupation functions in
different parts of the system are
overcome as well as the problem of how to normalize
the wavefunctions involved, since the Green's functions used in the
present 
description are always properly normalized. Our treatment is
adaptable to the inclusion of further many-body effects in the
problem, such as 
carrier-carrier interactions in the electrodes and the interaction
with phonon modes. 

One effect of biasing the system out of equilibrium is that 
the oscillatory exchange interaction in various dimensions exhibits
strong interference effects, leading to one or more changes of the
type of coupling (between FM and AFM) at a given impurity
configuration as the bias is varied. This behavior arises as a result
of an interference  between several fundamental oscillations due to
a mixing of different wavevectors. The possibility of tuning the
interaction 
through changing the bias alone
 could become an important effect in applications of nonlinear
switching devices using layered magnetic structures with a potential
barrier. 
Another important effect is that out of equilibrium the range of the
interaction increases, leading to an interaction energy that scales
extensively with the system size in the direction perpendicular to the
barrier for interacting slabs of spins. A closer study of
this phenomenon for the 3D case presented here, as well as for the 1D
case of interacting lines of spins, where the interaction in
equilibrium is known to lead to a helical ordering within each line,
is being undertaken.

Our results should be applicable to a very broad variety of
conceivable structures and the formalism we have presented here is
particularly 
suitable to be adapted to such situations. Extensions to include
varying effective masses and other material properties such as band
structure and different band filling in the various sub-parts of the
system are obvious. A particular example for a
possible extension would be to study the interaction in double or
multi-barrier systems out of equilibrium.
Furthermore, in order to facilitate a direct
comparison to experimental results  a next step could include the
calculation of the RKKY-perturbed spin-polarized tunneling current
across systems of this kind. We also hope our work will encourage the
experimental study of giant magnetoresistance
phenomena out of equilibrium where one can equally expect interesting
interference phenomena to occur as a result of the different relative
distances of the Fermi surfaces to the bottoms of the  conduction
bands involved.

\section{Acknowledgments}

The authors would like to thank Mr.\ Carsten Heide at Oxford for
reading the manuscript and making useful suggestions. One of the
authors (NFS) gratefully acknowledges financial support from NEC
for a visit to the NEC Research Institute at Princeton during this
collaboration. 

\appendix

\section{Nonequilibrium behavior of the interaction between slabs}

In the expression for the RKKY interaction between 3D slabs of
spins from (\ref{s9}), we exchange the spatial integrations with
the frequency integral, to obtain
\begin{eqnarray}
 & & E^{\rm s}_{\rm RK} = \frac{ \left( J \rho^{\rm 3D}_{s-d}
\right)^2 \nu^{\rm 2D}} {(2m)^4} {\bf S}_1 \cdot {\bf S}_2 
 \int_{-\infty}^{\infty} \frac {d  \omega} {2 \pi}
 \int \frac {d {\bf k}_{\|}^2}{(2 \pi)^2}  \int_{L-(d+l)}^{L-d} dx_1 
\int_{R+d}^{R+(d+l)} dx_2 \label{ap0} \\ 
& & {\rm Im} \left\{
\left[\eta_L(x_1)G_{(0)}(L,R)\eta_R(x_2) 
\right]^<  \left[ \eta_R^r(x_2)G_{(0)}^r(R,L)\eta_L^r(x_1)  +
\eta_R^a(x_2)G_{(0)}^a(R,L)\eta_L^a(x_1) \right] \right\} \nonumber.
\end{eqnarray}
In (\ref{ap0}) we have used (\ref{i9_5}) and (\ref{i9_6}) to rewrite
the 
expression
$G_{(0)}^<(x_1,x_2)$
$\left[G_{(0)}^r(x_2,x_1)+G_{(0)}^a(x_2,x_1)\right]$ 
in the 
integrand. In addition, we have introduced
\begin{eqnarray}
\eta_{L(R)}^{r/a}(x_{1(2)}) & = & \left. \partial_{x_0}
g^{r/a}_{L(R)}(x_{1(2)},x_0)\right\vert_{x_0=L(R)} =
\left. \partial_{x_0} 
g^{r/a}_{L(R)}(x_0,x_{1(2)})\right\vert_{x_0=L(R)} \nonumber \\
& = & + (-) 2m \exp\left\{\mp(\pm)iq_{L(R)}^{r/a}
[x_{1(2)}-L(R)]\right\} \label{ap1},
\end{eqnarray}
where $g^{r/a}_{L(R)}(x,x')$ is taken from (\ref{s1}).
Furthermore, the explicit frequency and wavevector
dependence of the integrand in (\ref{ap0}) was omitted for brevity.
The expression for
$G_{(0)}^<(x_1,x_2)\left[G_{(0)}^r(x_2,x_1)+G_{(0)}^a(x_2,x_1)\right]$
in the integrand 
can now be further grouped as follows:
\begin{eqnarray}
& &
G_{(0)}^<(x_1,x_2)\left[G_{(0)}^r(x_2,x_1)+G_{(0)}^a(x_2,x_1)\right] =
 \label{ap2} \\ 
& &   (2m)^{-4} \left\{ G_{(0)}^<(L,R)G_{(0)}^r(R,L)
[\eta_L^r(x_1)]^2\eta_R^r(x_2)\eta_R^a(x_2) \right. \nonumber \\ 
& + & G_{(0)}^<(L,R)G_{(0)}^a(R,L)
\eta_L^r(x_1)\eta_L^a(x_1)[\eta_R^a(x_2)]^2 \nonumber \\ 
& + & G_{(0)}^r(L,R)G_{(0)}^r(R,L)
[\eta_L^r(x_1)]^2\eta_R^r(x_2)\eta_R^<(x_2) \nonumber \\ 
& + & G_{(0)}^a(L,R)G_{(0)}^a(R,L)
\eta_L^<(x_1)\eta_L^a(x_1)[\eta_R^a(x_2)]^2 \nonumber \\ 
& + & G_{(0)}^r(L,R)G_{(0)}^a(R,L)
\eta_L^r(x_1)\eta_L^a(x_1)\eta_R^a(x_2)\eta_R^<(x_2)\nonumber \\ 
& + & \left. G_{(0)}^a(L,R) G_{(0)}^r(R,L)\eta_L^<(x_1)\eta_L^r(x_1)
\eta_R^r(x_2)\eta_R^a(x_2) \right\} . \nonumber
\end{eqnarray}
When now the functions $\eta^<$ are replaced by their definitions in
terms of retarded and advanced functions,
$\eta^<_{L(R)}=n_F^{L(R)}\left[ \eta^a_{L(R)}-\eta^r_{L(R)} \right]$,
one can see that all  
occurring $dx_1$ and $dx_2$ integrations can be accounted for by
introducing the following terms:
\begin{eqnarray}
\xi^0_{L(R)} & = & \int_{L-d-l}^{L-d}
 \left( \int_{R+d}^{R+d+l} \right ) dx_{1(2)} 
\eta_{L(R)}^r(x_{1(2)}) \eta_{L(R)}^a(x_{1(2)}) \label{ap3} \\
 & = & \frac {(2m)^2i} {q_{L(R)}^a-q_{L(R)}^r} \left\{ \exp \left[
i(q_{L(R)}^r-q_{L(R)}^a)(d+l) \right] - \exp \left[ 
i(q_{L(R)}^r-q_{L(R)}^a)d\right] \right\}
\nonumber \\
& = & \left\{ 
{ \begin{array}{cll}
(2m)^2 l &,  & \omega - V^{L(R)} \geq 0 \\
{\begin{array}{l}- \displaystyle{\frac {(2m)^2} {2 \sqrt{2m(V^{L(R)}-
\omega)}}} 
\left\{ \exp \left[ - 2 \sqrt{2m(V^{L(R)}- \omega)}(d+l) \right]
\right.  \\ 
- \left. \exp \left[ - 2 \sqrt{2m(V^{L(R)}- \omega)}d \right] \right\}
\end{array}} & ,  & 
 \omega - V^{L(R)} < 0
\end{array}} \right. , \nonumber \\ 
\xi_{L(R)}^{r/a} & = & \int_{L-d-l}^{L-d}
 \left( \int_{R+d}^{R+d+l} \right )  dx_{1(2)}
 [\eta_{L(R)}^{r/a}(x_{1(2)})]^2  \label{ap4} \\
 & = & \mp \frac {(2m)^2i}
 {2q_{L(R)}^{r/a}} \left\{ \exp \left[
\pm 2iq_{L(R)}^{r/a}(d+l) \right] - \exp \left[ \pm
2iq_{L(R)}^{r/a}d\right] \right\}.   \nonumber
\end{eqnarray}
While $\xi_{L(R)}^0$ from (\ref{ap3}) 
produces a term proportional to $l$ for energies above the 
band bottom in the respective lead, it is seen from (\ref{ap3}) and 
(\ref{ap4}) that
\begin{equation}
\xi_{L(R)}^0= \xi_{L(R)}^{r/a}, \hspace{5mm}  \omega - V^{L(R)} < 0.
\label{ap4_5}
\end{equation}
From here we find 
\begin{eqnarray}
& & \int_{L-(d+l)}^{L-d} dx_1 \int_{R+d}^{R+(d+l)} dx_2
G_{(0)}^<(x_1,x_2)\left[G_{(0)}^r(x_2,x_1)+G_{(0)}^a(x_2,x_1)\right] =
\label{ap5} \\ 
& & (2m)^{-4} \left\{ G_{(0)}^<(L,R)G_{(0)}^r(R,L) \xi^r_L \xi_R^0 +
G_{(0)}^<(L,R)G_{(0)}^a(R,L)\xi^a_R \xi_L^0 \right. \nonumber \\ 
& + & G_{(0)}^r(L,R)G_{(0)}^r(R,L) n^R_F(\omega) \xi^r_L
\left[\xi^0_R -\xi^r_R \right] +  
G_{(0)}^a(L,R)G_{(0)}^a(R,L) n_F^L(\omega) \xi^a_R \left[\xi^a_L -
\xi^0_L  \right]
\nonumber \\   
& + & \left. G_{(0)}^r(L,R)G_{(0)}^a(R,L) n_F^R(\omega) \xi^0_L
\left[\xi^a_R-\xi^0_R \right] + 
G_{(0)}^a(L,R)G_{(0)}^r(R,L) n_F^L(\omega) \xi^0_R \left[ \xi^0_L -
\xi^r_L 
\right]  \right\}. \nonumber
\end{eqnarray}
Since the problem we consider has time-reversal symmetry,
$G_{(0)}^{r/a}(L,R)=G_{(0)}^{r/a}(R,L)$ always holds.
 When (\ref{ap5}) is considered in
equilibrium, we know that also $G_{(0)}^<(L,R)=n_F(\omega) \left[
G_{(0)}^a(L,R)- 
G_{(0)}^r(L,R) \right]$ holds. In such a situation it is seen that
(\ref{ap5}) 
 simplifies to
\begin{eqnarray}
\int_{L-(d+l)}^{L-d} dx_1 \int_{R+d}^{R+(d+l)} dx_2
G_{(0)}^<(x_1,x_2)\left[G_{(0)}^r(x_2,x_1)+G_{(0)}^a(x_2,x_1)\right] =
\label{ap6} \\ 
n_F(\omega) \left\{ \left[ G_{(0)}^a(L,R) \right]^2 \xi_L^a \xi_R^a - 
 \left[ G_{(0)}^r(L,R) \right]^2  \xi_L^r \xi_R^r  \right\},
\nonumber 
\end{eqnarray}
where it was assumed that $\xi_L$ and $\xi_R$ may also be different in
equilibrium, e.g.\ if the materials on the left and right of the
junction have different work functions. Clearly in (\ref{ap6})
all factors proportional to $l$ have vanished as expected.
When $l$ is taken to $l \rightarrow \infty$
the functions $\xi^{r/a}_{L(R)}$ oscillate with $l$, as is seen from
(\ref{ap4}), and eventually reduce to 
\begin{equation}
\lim_{l \rightarrow \infty} \xi_{L(R)}^{r/a} = \pm \frac {(2m)^2i}
{2q_{L(R)}^{r/a}} \exp \left[ \pm 2iq_{L(R)}^{r/a}d\right],
\label{ap7} 
\end{equation}
due to the retarded/advanced property of the $q_{L(R)}^{r/a}$, which
causes an exponential decay in the relevant functions at large
distances. 

Out of equilibrium, however, the cancellation  of the
$l$-proportionality ceases to be complete and the expression for the
slab 
interaction (\ref{s9}) can be rewritten by means of (\ref{s5_9}) and
(\ref{n3}) for zero temperature as
\begin{eqnarray}
& & E^{\rm s}_{\rm RK} = \frac 
{\left( J \rho^{\rm 3D}_{s-d} \right)^2 \nu^{\rm 2D}}
{8\pi^2(2m)^3} {\bf S}_1  \cdot {\bf S}_2 \label{ap8} \\
& & \times \left\{ 
\int_0^{\mu} dz (\mu-z) {\rm Im} \left[
  \left(
\Gamma_L (z) - G_{(0)}^a(z) \right) \Xi(z) 
+ \left(G_{(0)}^a(z)\right)^2 \xi_L^a(z) \xi_R^a(z) \right] \right.
 \nonumber \\ 
& & + \left. \int_{-eV}^{\mu-eV} dz (\mu-z-eV) {\rm Im} \left[ \left( 
\Gamma_R (z) + G_{(0)}^r(z) \right) \Xi(z) -
\left(G_{(0)}^r(z)\right)^2 
 \xi_L^r(z) \xi_R^r(z) \right] \right\} \nonumber, 
\end{eqnarray}
where we have introduced
\begin{equation}
\Xi(z) = G_{(0)}^r(z) \xi_L^r(z)\xi_R^0(z) +
 G_{(0)}^a(z)\xi_R^a(z)\xi_L^0(z),  \label{ap9}
\end{equation}
with the abbreviation: $G_{(0)}(L,R;z)= G_{(0)}(z)$ and
$\Gamma_{L(R)}(L,R;z)= \Gamma_{L(R)}(z)$. Note  that in
(\ref{ap8}) terms containing $ \xi^0_L \xi^0_R G_{(0)}^r(z)
G_{(0)}^a(z)$ have 
vanished, since they are entirely real. From the definition of the 
$\xi^0_{L(R)}$ in  (\ref{ap3}) it is seen that the term $\Xi(z)$ from
(\ref{ap9}) is  
 proportional to $l$ for $z \geq 0$. When $-eV< z<0$, $\Xi(z)$ always
has one  
$l$-proportional and one exponentially decaying component, where the
latter 
describes the penetration of low-lying electron states in the right
lead  
to the position of the spins in the left one. The $l$-proportional
component, 
however, vanishes for energies below zero, since in this case the term
$\left[ \Gamma_R (z) + G_{(0)}^r(z) \right] G_{(0)}^r(z)
\xi_L^r(z)\xi_R^0(z)$ 
turns out to be entirely real.
The remaining $l$-dependence of $\Xi(z)$ subsequently translates to 
the slab interaction from (\ref{ap8}) for energies $z \geq 0$, 
where it is seen to produce terms
proportional to $l$, i.e.\ which scale extensively with the thickness
of the slabs, as well as the equilibrium terms from (\ref{ap6})
before, which 
show no $l$-proportionality.

\begin{figure}[tbp]

\caption{Schematic representation of the one dimensional tunneling
system described in the text. Two magnetic $s$-$d$ impurities are
situated in the left and right lead at equal distances $d$ away from
the electrode-barrier interfaces $L$ and $R$, respectively, and 
interact with each other through the tunneling of electrons across 
the barrier. The points $L$ and $R$ at the same time serve as the 
partitioning points within the Caroli/Feuchtwang formalism,
indicated by the two dashed vertical lines.}
\label{FIG1}
\end{figure}

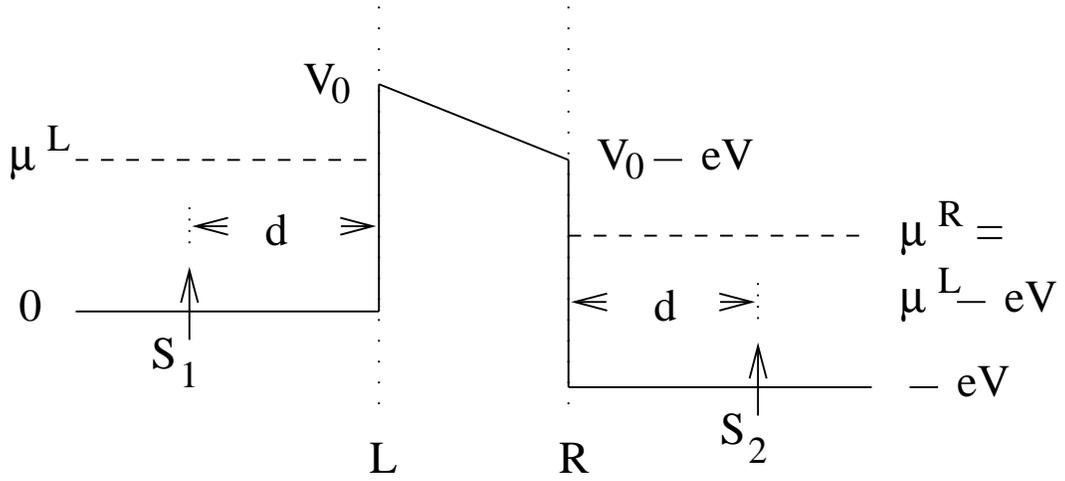
\begin{figure}[tbp]

\caption{Cross-sectional drawing of a magnetic multilayer structure,
obtained as a 3D planar extension of the 1D system shown
in Fig.\ 1. Two magnetic slabs M are separated by non-magnetic
spacer layers S and a planar tunneling barrier  B extending between
points $L$ and $R$.  The spins in the magnetic slabs are assumed to
have 
the same orientation within each slab due to a dominance of
ferromagnetic (FM) coupling at short distances.}
\label{FIG2}
\end{figure}

\begin{figure}[tbp]

\caption{Range functions: (a) $\Phi(x)$ for interacting magnetic
impurities in 1D from (59), (b)
$\Phi_{\rm m}(x)$ for interacting magnetic monolayers in 3D from (60),
and 
(c) $\Phi^{\infty}_{\rm s}(x)$ for semi-infinite interacting magnetic
slabs in 3D from (61) plotted against the distance 
$x=R-L$ between the spins in equilibrium, where the spins are
considered to be fixed on either interface of the barrier ($d=0$). The
barrier width is  
increased from $0.0 {\rm \AA}$ to $5.0 {\rm \AA}$ and the
height of the barrier is increased throughout the plots as
indicated. For $V_0/\mu=0.0$ (a), (b) and (c) show the range functions
$\Phi(x)$, $\Phi_{\rm m}(x)$ $\Phi^{\infty}_{\rm s}(x)$, respectively,
for a free electron system.}
\label{FIG3}
\end{figure}

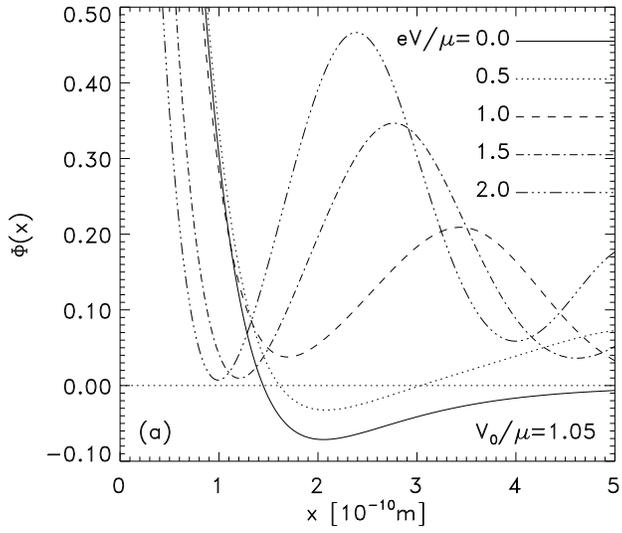
\begin{figure}[tbp]

\caption{(a),(b): range function $\Phi(x)$ from (59), (c),(d):
force function $-d \Phi(x)/dx$, and (e),(f): range function
$\Phi_{\rm m}(x)$ from (60) for finite bias $eV$.
The initial relative barrier height in equilibrium is
fixed at (a),(c),(e): $V_0/\mu=1.05$ and (b),(d),(f): $V_0/\mu=1.50$
with otherwise the same system parameters  
as in Fig.\ 3 ($\mu$ is the Fermi energy of the leads in equilibrium).
In four steps a bias $eV$ of up to $eV/\mu=2.0$ is applied to the
junction.}
\label{FIG4}
\end{figure}

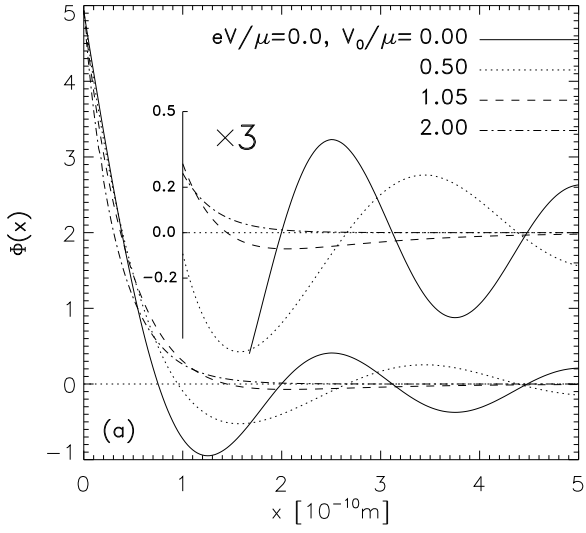
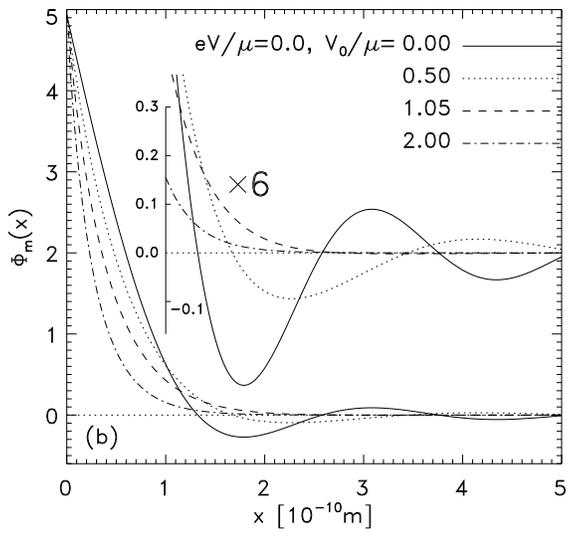
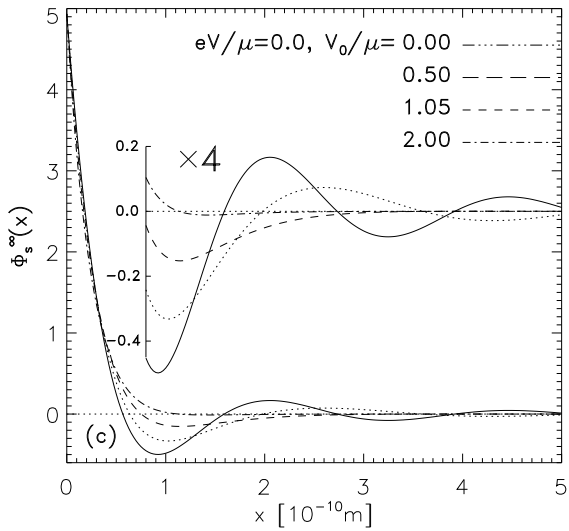
\begin{figure}[tbp]
\caption{ Surface plot of the range function (a): $\Phi$,  (b):
$\Phi_{\rm m}$ and (c): $\Phi'_s=[1+ 40 \times eV/\mu]^{-1}
\Phi^{l=10}_{\rm s}$ when the impurities are
moved within the electrodes. In the  present arrangement the barrier
height is fixed to $V_0/\mu=1.50$ and the barrier width to
$R-L=1.0 {\rm \AA}$. The distance $d$ of the impurities on either side
of the barrier is increased from $d=(0.0-5.0) {\rm \AA}$ (plotted
across), while at the same time the bias is varied from
$eV=(0.0-1.5){\rm eV}$ (plotted into depth). The thickness of the
slabs in (c) was taken to be $l=10{\rm \AA}$. In (a), (b) and (c)
the front faces of the surfaces show
the equilibrium range functions, which are oscillatory as the spins
are moved through the leads. In order to be able to better
identify the boundaries between FM and AFM regions, we have
overlayed a contour plot showing the zero coupling ($\Phi=0$) 
contour.}
\label{FIG5}
\end{figure}

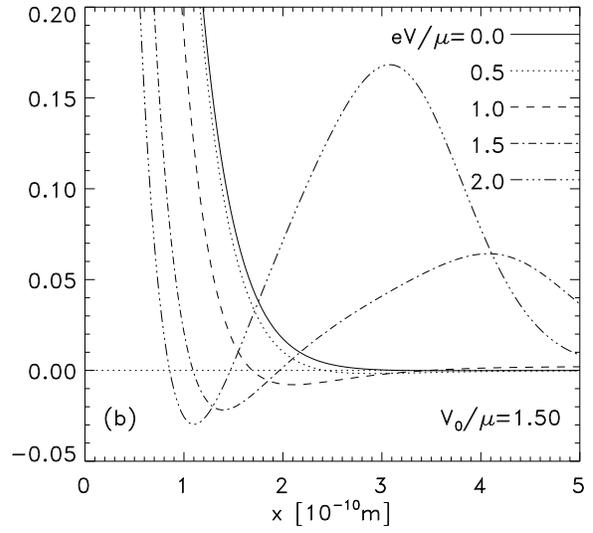
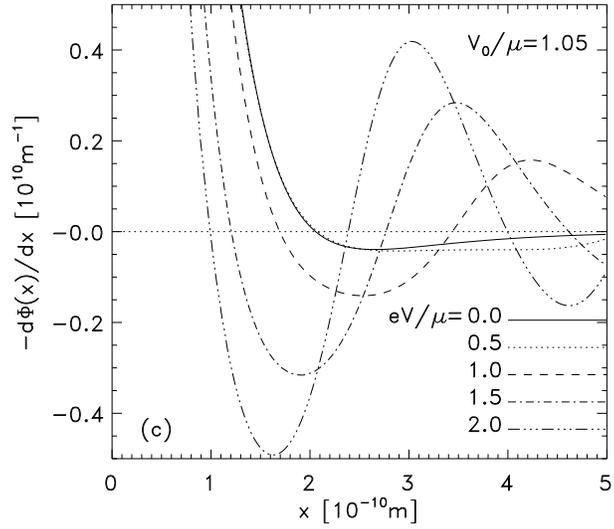
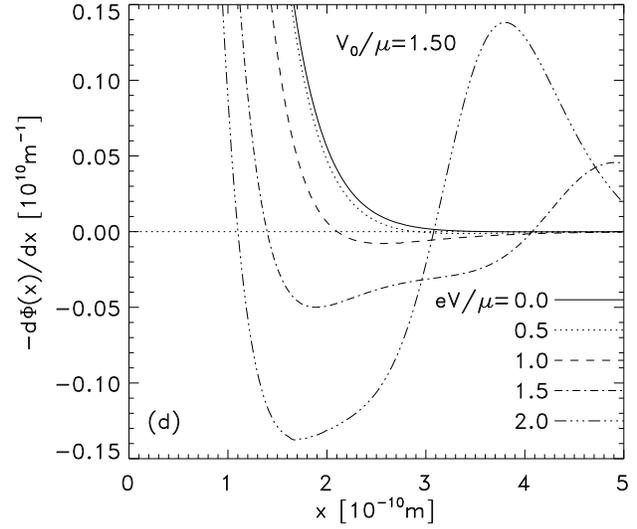
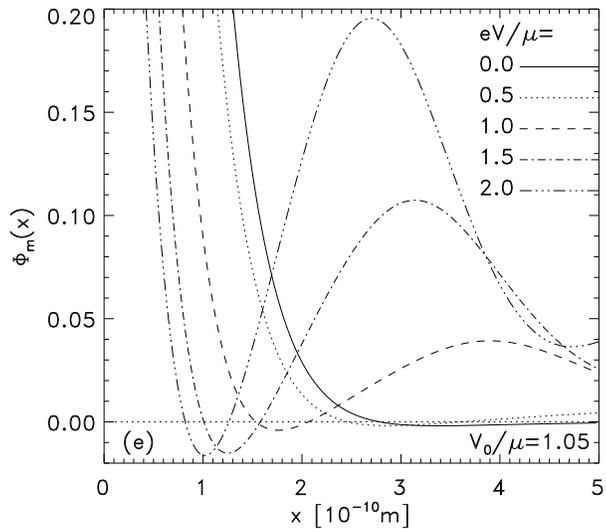
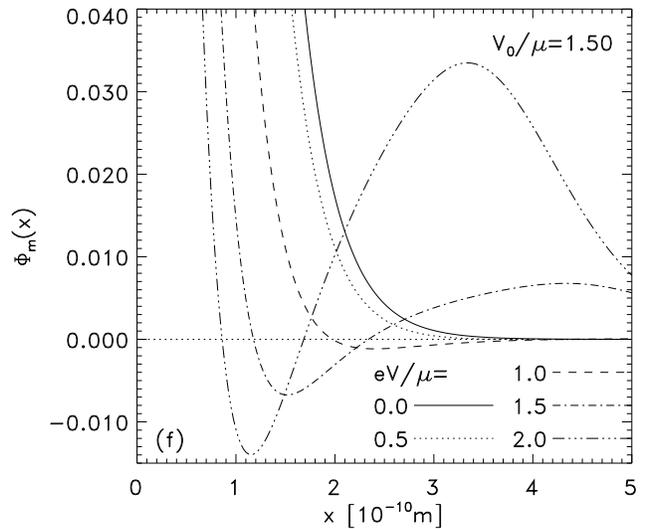
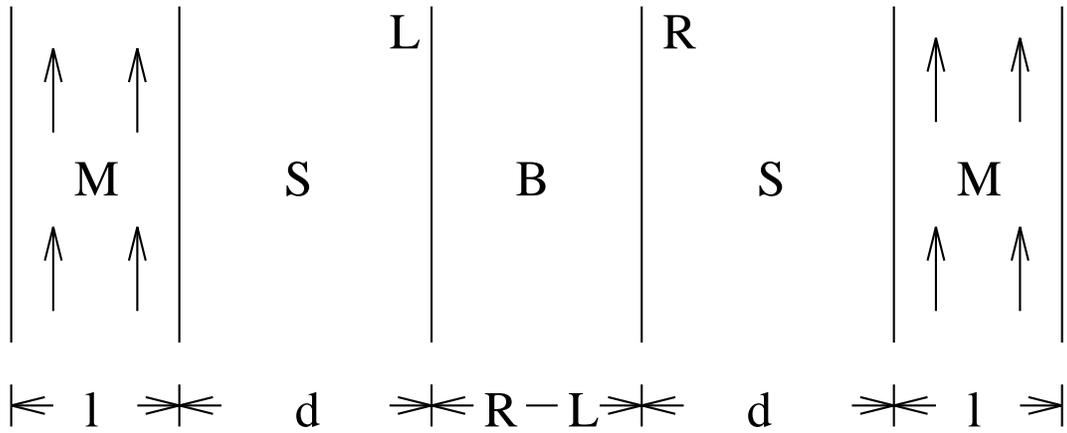
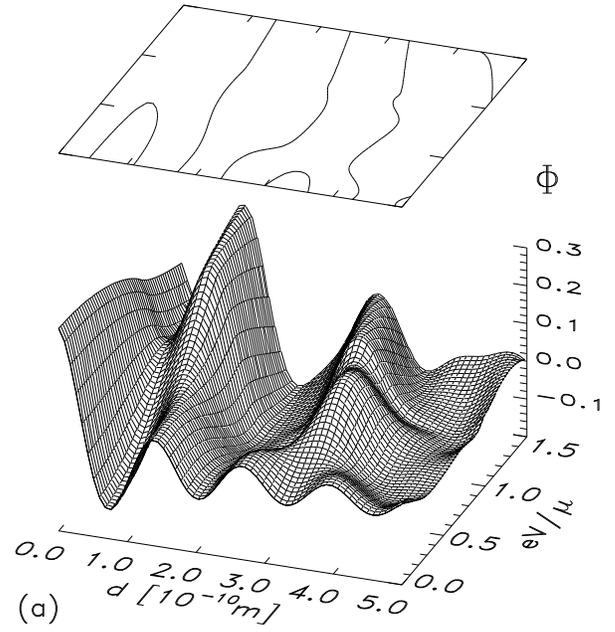
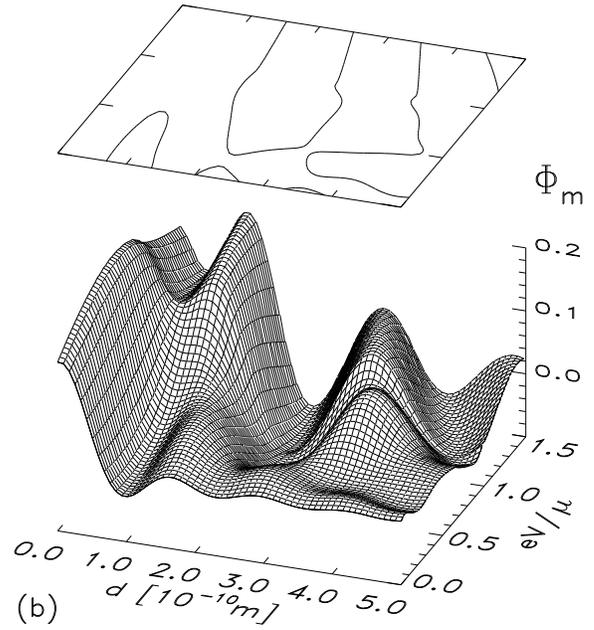
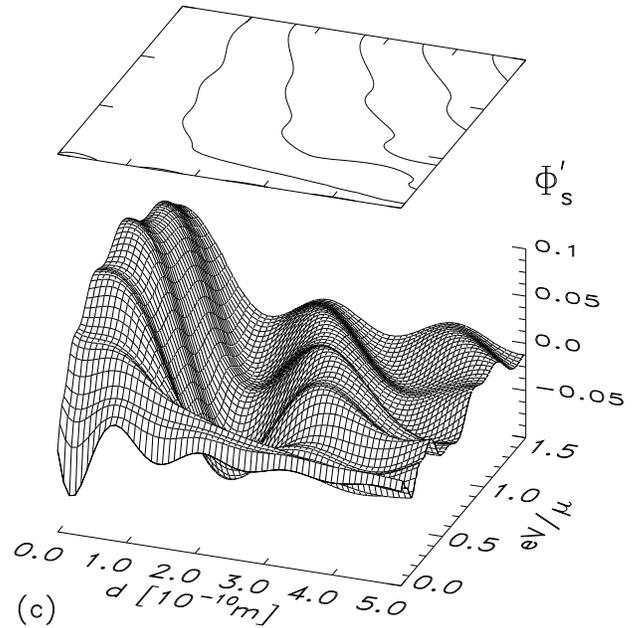
\end{document}